\documentclass[twocolumn, apj, iop]{aastex62}

\usepackage{amsmath}
\usepackage{natbib}
\usepackage{gensymb}
\usepackage{graphicx}
\usepackage[utf8x]{inputenc}

\graphicspath{./}

\submitjournal{ApJ}
\accepted{2019/9/24}

\shorttitle{IS THE AGN FEEDBACK CYCLE BROKEN IN A2495?}
\shortauthors{Pasini et al.}

\begin{document}

\title{A BCG WITH OFFSET COOLING: IS THE AGN FEEDBACK CYCLE BROKEN IN A2495?}

\correspondingauthor{Myriam Gitti}
\email{myriam.gitti@unibo.it}

\author{T. Pasini}
\affiliation{Dipartimento di Fisica e Astronomia (DIFA), Universita` di Bologna, via Gobetti 93/2, 40129 Bologna, Italy}

\author[0000-0002-0843-3009]{M. Gitti}
\affiliation{Dipartimento di Fisica e Astronomia (DIFA), Universita` di Bologna, via Gobetti 93/2, 40129 Bologna, Italy}
\affiliation{Istituto Nazionale di Astrofisica (INAF) – Istituto di Radioastronomia (IRA), via Gobetti 101, I-40129 Bologna, Italy}

\author{F. Brighenti}
\affiliation{Dipartimento di Fisica e Astronomia (DIFA), Universita` di Bologna, via Gobetti 93/2, 40129 Bologna, Italy}

\author[0000-0002-8341-342X]{P. Temi}
\affiliation{Astrophysics Branch, NASA/Ames Research Center, MS 245-6, Moffett Field, CA 94035}

\author{A. Amblard}
\affiliation{Astrophysics Branch, NASA/Ames Research Center, MS 245-6, Moffett Field, CA 94035}
\affiliation{BAER Institute, Sonoma, CA, USA}

\author{S. L. Hamer}
\affiliation{Department of Physics, University of Bath, Claverton Down, BA2 7AY, UK}

\author{S. Ettori}
\affiliation{Istituto Nazionale di Astrofisica (INAF) – Osservatorio di Astrofisica e Scienza dello Spazio (OAS), via Gobetti 93/3, I-40129 Bologna, Italy}
\affiliation{Istituto Nazionale di Fisica Nucleare (INFN) – Sezione di Bologna, viale Berti Pichat 6/2, I-40127 Bologna, Italy}

\author[0000-0002-5671-6900]{E. O'Sullivan}
\affiliation{Harvard-Smithsonian Center for Astrophysics, 60 Garden Street, Cambridge, MA02138, USA}

\author[0000-0002-9112-0184]{F. Gastaldello}
\affiliation{INAF-IASF Milano, via E. Bassini 15, I-20133 Milano, Italy}

\begin{abstract}

We present a combined radio/X-ray analysis of the poorly studied galaxy cluster Abell 2495 (z=0.07923) based on new EVLA and \textit{Chandra} data. We also analyze and discuss H$\alpha$ emission and optical continuum data retrieved from the literature. We find an offset of $\sim$ 6 kpc between the cluster BCG (MCG+02-58-021) and the peak of the X-ray emission, suggesting that the cooling process is not taking place on the central galaxy nucleus. We propose that sloshing of the ICM could be responsible for this separation. Furthermore, we detect a second, $\sim$ 4 kpc offset between the peak of the H$\alpha$ emission and that of the X-ray emission. Optical images highlight the presence of a dust filament extending up to $\sim$ 6 kpc in the cluster BCG, and allow us to estimate a dust mass within the central 7 kpc of 1.7 $\cdot$ 10$^5$ M$_\sun$. Exploiting the dust to gas ratio and the $L_{\text{H}\alpha}$-$M_{\text{mol}}$ relation, we argue that a significant amount (up to 10$^9$ M$_\sun$) of molecular gas should be present in the BCG of this cluster. We also investigate the presence of ICM depressions, finding two putative systems of cavities; the inner pair is characterized by $t_{\text{age}} \sim 18$ Myr and $P_{\text{cav}} \sim$ 1.2 $\cdot$ 10$^{43}$ erg s$^{-1}$, the outer one by $t_{\text{age}} \sim 53$ Myr and $P_{\text{cav}} \sim$ 5.6 $\cdot$ 10$^{42}$ erg s$^{-1}$. Their age difference appears to be consistent with the free-fall time of the central cooling gas and with the offset timescale estimated with the H$\alpha$ kinematic data, suggesting that sloshing is likely playing a key role in this environment. Furthermore, the cavities' power analysis shows that the AGN energy injection is able to sustain the feedback cycle, despite cooling being offset from the BCG nucleus. 

\end{abstract}


\section{Introduction} \label{sec:intro}

The classical cooling flow model predicted that the intra-cluster
medium (ICM) of cool-core galaxy clusters should cool, condense and
accrete onto the brightest cluster galaxy (BCG), forming stars and
producing strong line emission radiation \citep[][]{Fabian_1994}. These features
are seen in many galaxy clusters, but both at a rate $\sim 1-10$ $\%$
of that expected in the standard model \citep[][]{Peterson-Fabian_2006}.
\\
\indent It is now widely recognized that cooling of the central gas is
quenched by Active Galactic Nuclei (AGN) found in the BCGs of
clusters \citep[][]{McNamara-Nulsen_2007, Gitti_2012, Fabian_2012}. In
this "radio-mode" mechanical feedback, radio jets or outflows
inflate bubbles (seen as X-ray brightness depressions named
\textit{cavities}) and generate shock waves and cold fronts in the hot
atmosphere of the cluster \citep[][]{McNamara_2000, Fabian_2006}. The
AGN is fueled through Super Massive Black Hole (SMBH) accretion of the
same gas condensing from the central regions \citep[][]{Gaspari_2011a,
  Gaspari_2013, Soker-Pizzolato_2005}, establishing a feedback loop in
which the various components are able to regulate each other. This
hypothesis is supported by the correlation between the AGN mechanical power,
estimated from the X-ray cavities, and the ICM cooling rate
\citep[e.g.,][]{Birzan_2004, Rafferty_2006}.
\\ \indent
BCGs at the center of galaxy clusters often host a rich multiphase
medium which can extend for tens of kpc, as revealed by strong line
emission from ionized (warm) and molecular (cold) gas
\citep[][and references therein]{Crawford_1999, Hamer_2016, McDonald_2010, McDonald_2014,
 Russell_2019}. The warm and cold gas show correlations with each
other and with the hot ICM \citep[][]{Crawford_1999, Edge_2001, Hogan_2017, Pulido_2018}, which strongly suggest that hot gas
cooling (albeit reduced with respect to the classical cooling flow
model) is the origin of the observed cold gas. Observations, simulations 
and analytic investigations agree that spatially extended cooling is likely to occur
in dense cool cores with short cooling times \citep[or cooling time/dynamical time 
ratio below certain threshold, e.g.,][and references therein]{Hogan_2017, Pulido_2018}.
The multiphase medium in cluster cores also reveals a complex
dynamics, likely the result of AGN activity and merging events. 
Chaotic (turbulent) motion and outflows are common \citep{Heckman_1989,
Hamer_2016, Russell_2019}.
It seems reasonable that cold gas inherits the disturbed
dynamics from the (low entropy) hot gas from which it has cooled
\citep[][]{McDonald_2010, McNamara_2016, Gaspari_2018}.
\\ \indent
In dynamically-relaxed systems, we expect the BCG to be at the centre of the cluster potential well, as described by  the 'central galaxy paradigm' \citep[][]{VandenBosch_2005, Cui_2016}. In this scenario, the galaxy should be coincident both with the cluster cool core centre (i.e. the X-ray peak) and with the line emission peak. However, in the case of interactions with other clusters or halos, all these components are likely to shift, leading to the production of offsets between them.  The connection between offsets and the dynamical state of clusters has been investigated both by observational studies \citep[][]{Katayama_2003, Patel_2006, Rossetti_2016}, making use of the current generation of X-ray satellites (e.g., \textit{Chandra} and \textit{XMM-Newton}),  and simulations \citep[]{Skibba_2011}.  \citet{Hudson_2010} studied a sample of 64 cool-core (CC) and non cool-core (NCC) clusters, finding that objects with a projected distance $> 50$ h$^{-1}_{71}$ kpc between the BCG and the X-ray peak (12\% of the sample) typically show large-scale radio emission, that is believed to be associated with major mergers; these clusters are usually NCC. On the other hand, the vast majority (80\%) of CC present an offset $< 50$ h$^{-1}_{71}$ kpc; among these, only 2 clusters show large-scale radio emission. A similar study by \citet{Sanderson_2009} on a sample of 65 CC clusters found that stronger cool cores are associated with smaller offsets, since more dynamically-disturbed objects have likely undergone a stronger merging phase that produced a disruptive impact on the X-ray core. The same trend was confirmed by \citet{Mittal_2009}. Offsets of the line emission peak are significant for the comprehension of the clusters dynamical evolution, too: detecting 10$^{4}$ K gas shifted from the BCG \citep[e.g.,][]{Sharma_2004} could suggest that cooling of the ICM is able to reach low temperatures even outside the central galaxy environment \citep{Hamer_2012}. \added{In addition, offsets between the H$\alpha$ and the X-ray peaks were found in A1795 by \citet{Crawford_2005} and in several clusters studied in the work by \citet{Hamer_2016}.}
\\ \indent
These works have established the correlation between offsets and clusters dynamical state \citep[e.g.,][]{Rossetti_2016}, which can therefore be used in order to discriminate between dynamically relaxed and non-relaxed objects \citep[e.g.,][]{Sanderson_2009, Hudson_2010, Mann_2012}.
\\ 
\\ 
This work is a multi-frequency study of Abell 2495 (A2495). This object was selected from the ROSAT BCS Sample \citep[][]{Ebeling_1998} by choosing objects with X-ray fluxes greater than $10^{-11} $ erg  cm$^{-2}$ s$^{-1}$ (1.18 $\cdot 10^{-11}$ erg cm$^{-2}$ s$^{-1}$ for A2495) and, among these, by selecting those characterized by L$_{H\alpha} >$ 10$^{40}$ erg s$^{-1}$ from the catalogue of \citet{Crawford_1999}. Following these criteria, we obtained a compilation of 13 objects, including some of the best-studied galaxy clusters (e.g., A1795, A478); among all these objects, A2495 and A1668 were still unobserved by \textit{Chandra}. We thus proposed joint \textit{Chandra}/VLA observations (P.I. Gitti) of these two clusters and were awarded time in \textit{Chandra} cycles 11 and 12. This paper presents our results for A2495, and a forthcoming paper will describe A1668 (Pasini et al., in preparation). A2495 was previously observed by NVSS, which
provides an estimate of the 1.4 GHz flux of \textit{F}(1.4 GHz) = 14.7
$\pm$ 0.6 mJy, and by TGSS, from which \textit{F}(150 MHz) $\simeq$
136 mJy. We also discuss H$\alpha$ data from \citet{Hamer_2016} and
exploit optical images (P.I. Zaritsky) from the \textit{Hubble Space Telescope}
(HST) archive. With this broad data coverage we are in the position to
perform a multi-wavelength investigation of this cluster, that will allow us to better understand the dynamical interactions between the hot ICM, the colder gas and the radio galaxy hosted in the BCG.
\\ \indent
We adopt a $\Lambda$CDM cosmology with $H_0$ = 73 km s$^{-1}$ Mpc$^{-1}$, $\Omega_M $= $ 1-\Omega_\Lambda $ = 0.3, and assumed the BCG redshift $z$ = 0.07923 \citep{Rines_2016, Hamer_2016} for the cluster as a whole. The luminosity distance is 345 Mpc, leading to a conversion of 1$''$ = 1.44 kpc.

\section{Radio analysis}

\subsection{Observations and data reduction}
\label{sec:radio}

We performed new observations of the radio source associated with A2495 BCG ($RA$ = 22$^\text{h}$50$^\text{m}$19.7$^\text{s}$, $DEC$ = +10$\degree$54$^\text{m}$12.7$^\text{s}$, J2000) at 5 GHz and 1.4 GHz with the EVLA in, respectively, B and A configuration (for details, see Table \ref{tab:radiobs}).

\begin{table*}[!htb]
	\centering 
	\begin{tabular}{c c c c c c c c}
		\hline
		\hline
		Data &  P.I. & Project Code & Number of spw & Channels & Bandwith & Array & Total exposure time \\
		\hline
		5 GHz (C BAND) & M. Gitti & SC0143 & 2 (4832 MHz - 4960 MHz) & 64 & 128 MHz & B & 7h58m35s \\
		1.4 GHz (L BAND) &  M. Gitti & SC0143 & 2 (1264 MHz - 1392 MHz) & 64 & 128 MHz &A & 3h59m22s \\
		\hline
	\end{tabular}
	\caption{Radio observations properties.} \label{tab:radiobs}
\end{table*}

 For both the observations the source 3C48 (J0137+3309) was used as primary flux calibrator; for the 5 GHz observation, that was split into two datasets, J2241+0953 was used as secondary phase calibrator and 3C48 as polarization calibrator; for the 1.4 GHz observation, J2330+1100 was used as both secondary phase and polarization calibrator.
\\ \indent
The data reduction was performed with the NRAO Common Astronomy Software Applications package (CASA), version 5.3. We applied the standard calibration procedure on each dataset and carried out an accurate editing of the visibilities. We performed both manual and automatic flagging (task {\ttfamily FLAGDATA}, mode= manual and rflag) to exclude radio frequency interferences (RFI) and corrupted data. As a result, we removed about 10\% of the visibilities in the 5 GHz band and about 20\% in the 1.4 GHz band. We also attempted the self-calibration on the target, but the operation was not successful probably due to the source faintness.
\\ \indent
We then applied the standard imaging procedure, using the {\ttfamily CLEAN} task, on a 7$''$ $\times$ 7$''$ region centered on the cluster. We made use of the {\ttfamily gridmode=WIDEFIELD} option in order to parametrize the sky curvature; we also set a two-terms approximation of the spectral model by using {\ttfamily nterms=2} (MS-MFS algorithm, \citet{Rau_2011}) and performed a multi-scale clean ({\ttfamily multiscale=[0,5,24]}), in order to better reconstruct the faint extended emission.

\subsection{Results}
\label{cap:results}

For each observing band, we produced maps by setting {\ttfamily weighting=BRIGGS} and {\ttfamily ROBUST 0}, in order to obtain the best combination of resolution and sensitivity. The {\ttfamily NATURAL} and {\ttfamily UNIFORM} maps, that  respectively enhance the sensitivity and the resolution, did not exhibit any interesting feature compared to the {\ttfamily ROBUST 0} image, so they are not shown here.  Typical amplitude calibration errors of our data were at 8$\%$: this is the uncertainty we assumed for the flux density measurements.

\subsubsection{5 GHz map}

\begin{figure}[h]
	\centering
	\includegraphics[width=26em]{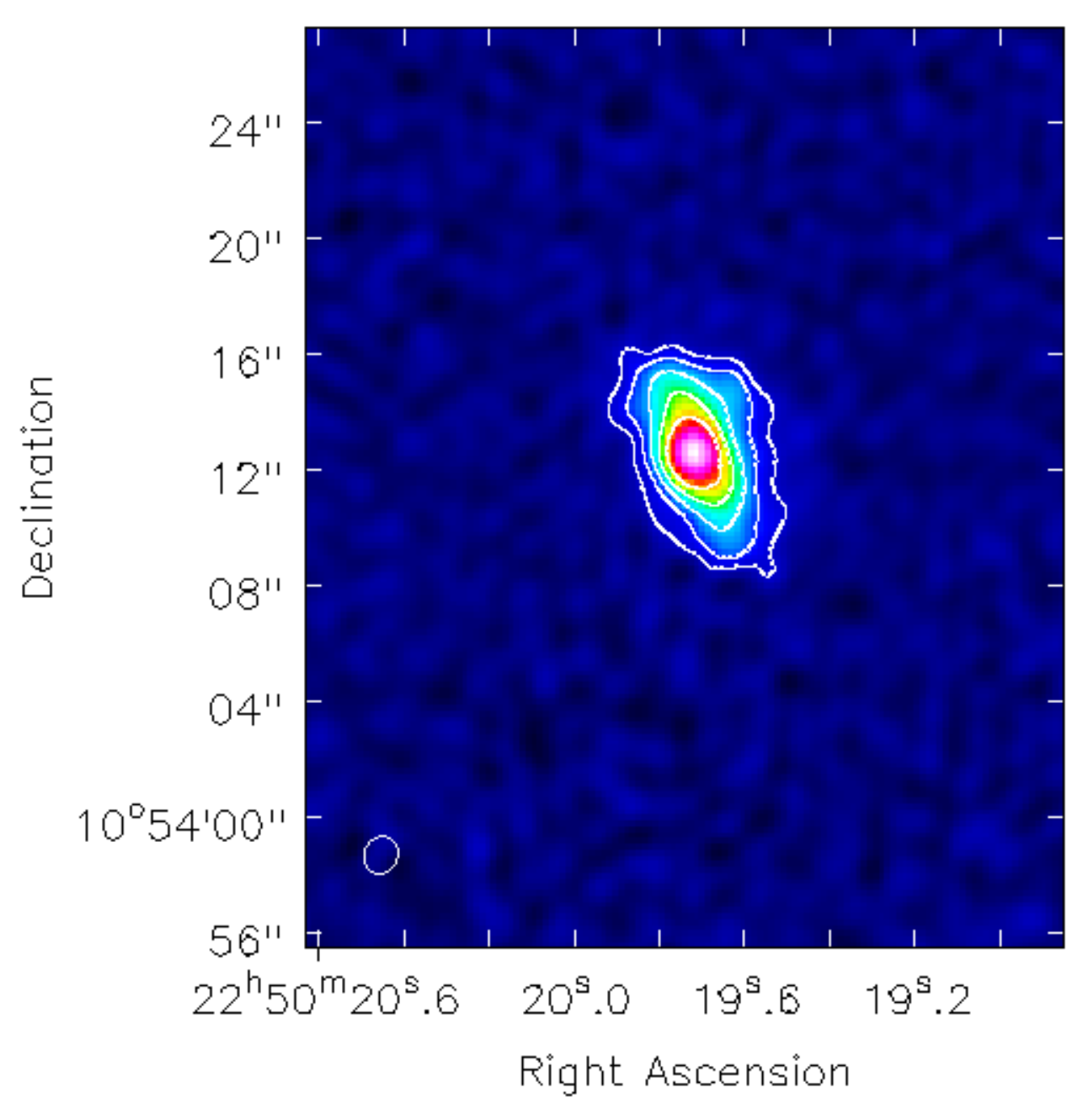}
	\caption{5 GHz map ({\ttfamily ROBUST 0}) of A2495. The resolution is 1.30$''$ $\times$ 1.15$''$, with a RMS noise of 4 $\mu$Jy beam$^{-1}$. Countours are at -3,3,6,12,24,48 $\cdot$ RMS. The source flux is 2.37 $\pm$ 0.19 mJy. The bottom-left circle represents the beam.}
	\label{fig:Cbriggs}
\end{figure}

The 5 GHz map (Fig. \ref{fig:Cbriggs}) is characterized by a resolution of 1.30$''$ $\times$ 1.15$''$ and a RMS noise of $ 4 \ \mu$Jy \ beam$^{-1}$. The total 5 GHz flux of the radio source in A2495 is 2.37 $\pm$ 0.19 mJy. This result is slightly different from \citet{Hogan_2015}, who measured a source flux of 2.85 $\pm$ 0.09 mJy in C array. This difference can be produced either by observational or intrinsic effects:  the maximum baseline of C configuration, used by \citet{Hogan_2015}, is shorter with respect to the B array, therefore its brightness sensitivity is higher; on the other hand, short-timescale (weeks to months) 10-30\% flux variations in radiogalaxies have recently been observed, especially in cool core cluster's BCGs \citep[e.g.,][]{Dutson_2014, Hogan_2015b}.
\\ \indent
We find that the radio emission is produced by the central radio galaxy on a scale of $\sim$ 10$''$; we do not observe any feature of diffuse emission. Following the method described by \citet{Feretti-Giovannini_2008}, we estimated the equipartition field, finding $B_{eq}$(5 GHz) = 3.81 $\pm$ 0.05 $\mu$G, consistent with the typical values characterizing radio galaxies.
\\ \indent
We used {\ttfamily stokes=Q} and {\ttfamily stokes=U} maps in order to get informations about the radio source polarization, and found that the polarization percentage at 5 GHz is about 5$\%$. 
Table \ref{tab:properties} summarizes the radio properties of the source on each band. 

\begin{table*}[!htb]
	\centering 
	\begin{tabular}{c c c c c c c}
		\hline
		\hline
		Band & Flux  & Luminosity$^{(a)}$ & Volume & Brightness Temperature & Equipartition Field & Polarisation \\
		&  [mJy] & [10$^{22}$ W Hz$^{-1}$] & [kpc$^3$] & [K] & [$\mu$G] & \\
		\hline
		5 GHz & 2.37 $\pm$ 0.19 &  3.3 $\pm$ 0.3 & 616 $\pm$ 56  & 3.7 $\pm$ 0.9 & 3.81 $\pm$ 0.05 & 5$\%$ \\
		1.4 GHz & 15.70 $\pm$ 1.30  & 21.8 $\pm$ 1.8  & 1112 $\pm$ 78 & 232 $\pm$ 49 & 5.60 $\pm$ 0.07 & $\leq$ 1$\%$ \\ 
		\hline
	\end{tabular}
	\caption{Radio properties of A2495. The volume is estimated assuming a prolate elissoid shape, where the axes are \textit{a} = 13.6 $\pm$ 1.8 kpc, \textit{b} = 9.3 $\pm$ 1.8 kpc for the 5 GHz map and \textit{a} = 13.6 $\pm$ 1.8 kpc, \textit{b} = 12.5 $\pm$ 1.8 kpc for the 1.4 GHz map. \\ ${(a)}$: estimated using L$_\nu$ = $4\pi \cdot (D_{L})^2 \cdot F_\nu \cdot (1+z)^{\alpha -1}$, where $D_{L}$ is the luminosity distance, $F_\nu$ is the flux at the frequency $\nu$ and $\alpha$ is the mean spectral index (see Sec. \ref{sec:spindex}).}
	\label{tab:properties}
\end{table*}

\subsubsection{1.4 GHz map}

\begin{figure}[h]
	\centering
	\includegraphics[width=24em, height=25.5em]{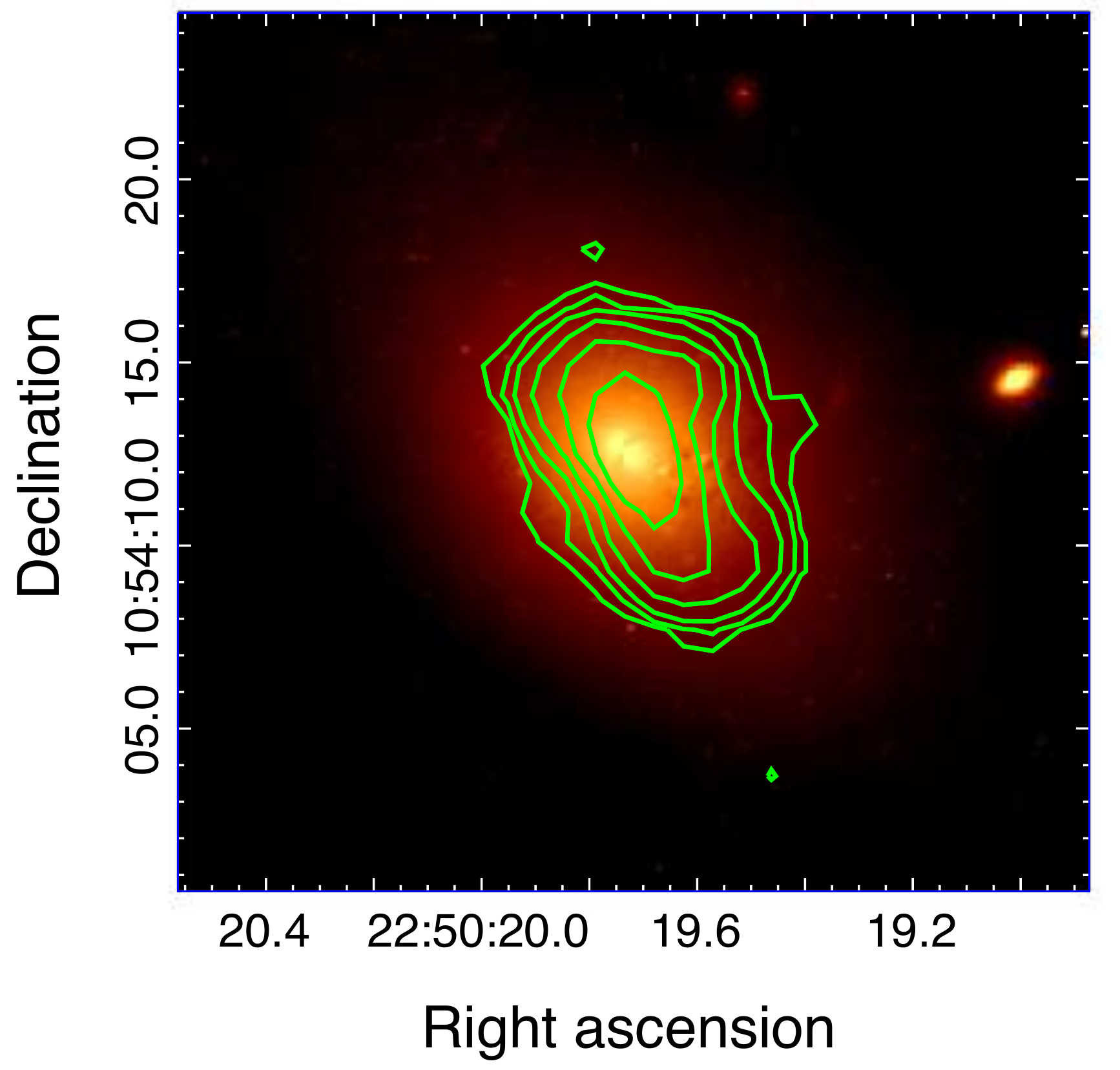}
	\caption{1.4 GHz contours ({\ttfamily ROBUST 0}) of the radio source overlaid on an HST image (F606W filter) of A2495. The resolution of the radio map is 1.29$''$ $\times$ 1.12$''$, with a RMS noise of 10 $\mu$Jy beam$^{-1}$. Countours are at -3,3,6,12,24,48 $\cdot$ RMS. The source flux is 15.7$\pm$ 1.3 mJy.} 
	\label{fig:Lbriggs}
\end{figure}

In Fig. \ref{fig:Lbriggs} we overlaid the 1.4 GHz contours on an optical image (F606W filter) retrieved from the HST Archive. 
The total source flux at this frequency is 15.7 $\pm$ 1.3 mJy, in agreement with the estimation made by \citet{Owen_1997} of $\sim$ 14 mJy. The radio source emission entirely lies on the BCG, whose major axis measures $\sim$ 25 kpc; there are no apparent hints of emission on larger scales. This is confirmed by the lack of additional flux on short baselines and by the flux estimate given by NVSS (see Section \ref{sec:intro}), consistent with ours. The radio galaxy morphology is very similar to the one at 5 GHz, extending on a scale of about $\sim$ 10$''$ ($\sim$ 14 kpc). Radio properties at 1.4 GHz can be found in Table \ref{tab:properties}.
\\ \indent
The 1.4 GHz luminosity is 2.18 $\cdot$ 10$^{23}$ W Hz$^{-1}$: the radio source in A2495 can be classified as FRI galaxy, characterised by asymmetric lobes and absence of hotspots. Similarly to what done at 5 GHz, we estimated the equipartition field, finding 5.60 $\pm$ 0.07 $\mu$G, and determined an upper limit of $\sim$ 1$\%$ for the polarization.
\\ \indent
\citet{Hogan_2015} performed a study of the radio properties of a large sample of BCGs, producing the luminosity function at 1.4 GHz of the BCS Sample. Notably, our luminosity estimate places A2495 in the 80$^{\text{th}}$ percentile of the function (see Fig.4 of \citealt{Hogan_2015}); of the 13 clusters that meet our selection criteria, it is the least radio powerful.

\subsubsection{Spectral index map}
\label{sec:spindex}

\begin{figure}[h]
	\centering
	\includegraphics[width=29em]{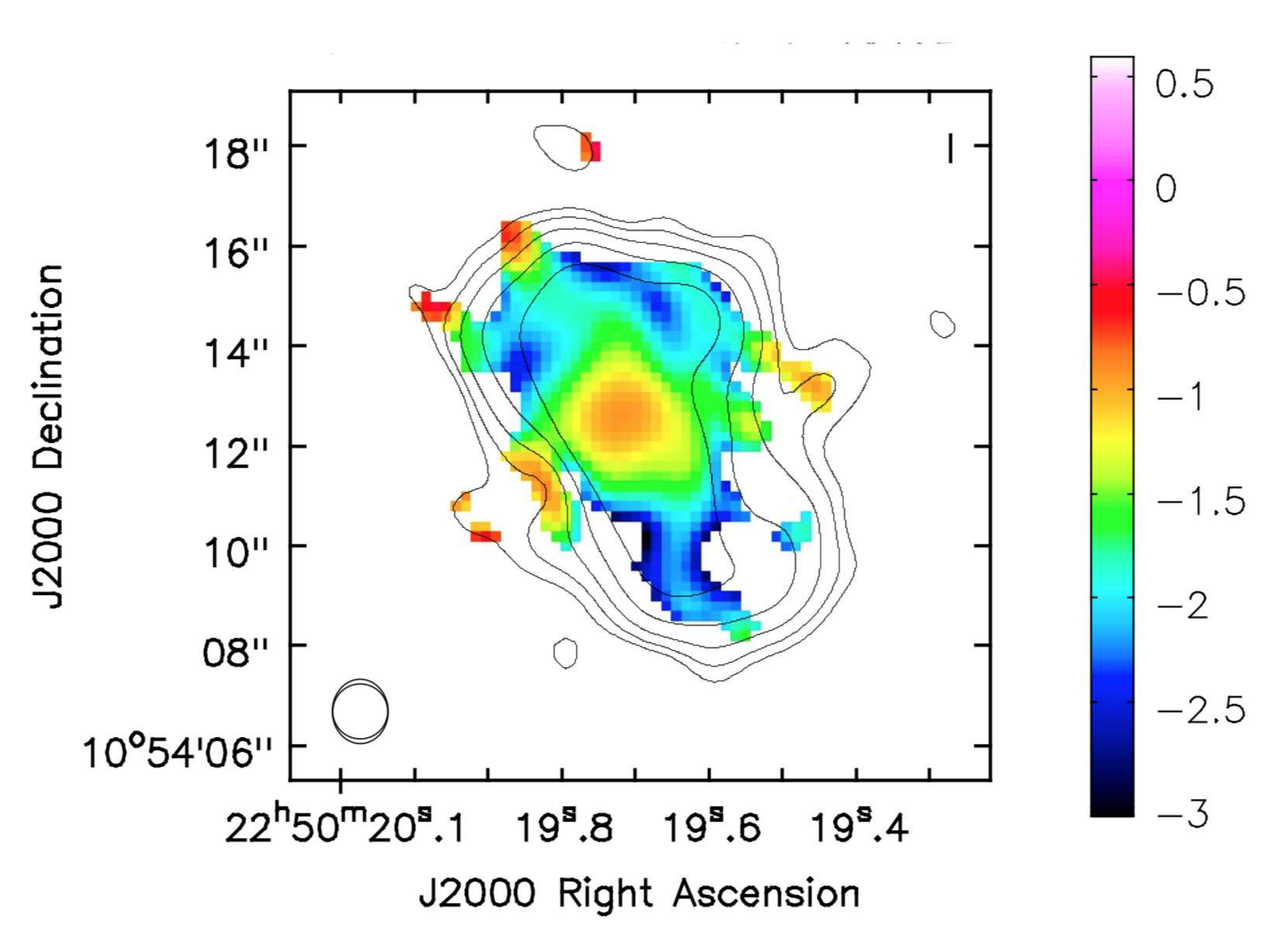}
	\caption{Spectral index map of A2495 between 5 GHz and 1.4 GHz. Countours are the same as the 1.4 GHz map, and typical errors range from $\Delta \alpha \simeq$ 0.2 to $\Delta \alpha \simeq$ 0.6 for the inner and outer regions, respectively.}
	\label{fig:spix}
\end{figure}

We produced the spectral index map, via the CASA task {\ttfamily IMMATH}, by combining the 1.4 GHz and the 5 GHz ones produced by setting {\ttfamily weighting=UNIFORM, uvrange=6-178} (in order to match uvranges between the two bands), with a resolution of 1.1$''$ $\times$ 1.1$''$. The result is shown in Fig \ref{fig:spix}.

The synchrotron index $\alpha$ is defined as:

 \begin{equation}
 \alpha =  - \dfrac{log \frac{S_1}{S_2}}{log \frac{\nu_1}{\nu_2}}
 \end{equation}

where, in this work, $S_1$ and $S_2$ are, respectively the 5 GHz flux (hereafter $S_C$) and the 1.4 GHz flux (hereafter $S_L$), while $\nu_1$ and $\nu_2$ are the corresponding frequences.
The radio galaxy core, often characterised by $\alpha \sim$ - 0.5, exhibits $\alpha \simeq$ - 0.9, while lobes reach $\alpha \simeq$ - 2. Table \ref{tab:spindex} summarizes the spectral index properties.

\begin{table}[!htb]
	\centering 
	\begin{tabular}{c c c c }
		\hline
		\hline
		Region & $S_C \pm \Delta S_C$ & $S_L \pm \Delta S_L$ & $\alpha$ $\pm$ $\Delta \alpha$\\
		& [mJy] & [mJy] & \\
		\hline
		Peak & 0.77 $\pm$ 0.06 & 2.20 $\pm$ 0.18 & -0.79 $\pm$ 0.08\\
		Extended$^{(a)}$ & 1.60 $\pm$ 0.16 & 13.50 $\pm$ 1.08 & -1.59 $\pm$0.08\\
		Total & 2.37 $\pm$ 0.19 & 15.70 $\pm$ 1.26 & -1.39 $\pm$ 0.22\\
		\hline
	\end{tabular}
	\caption{The first column shows the flux values at 5 GHz, while the second displays the 1.4 GHz values. The third column presents the correspondent spectral index values.\\ \textit{(a)}: estimated subtracting the peak contribute from the total flux.} \label{tab:spindex}
\end{table}

The mean index is - 1.39 $\pm$ 0.22, in agreement with \citet{Hogan_2015}, that determined $\alpha \simeq$ - 1.35, thus suggesting the presence of an old electronic population.

\section{X-ray Analysis}

\subsection{Observation and data reduction}

A2495 has been observed with the \textit{Chandra Advanced CCD Imaging Spectrometer S} (ACIS-S) on 17/07/2012 (ObsID 12876, P.I. Gitti) for a total exposure of $\sim$ 8 ks. 
Data were reprocessed with CIAO 4.9 using CALDB 4.2.1. First, the {\ttfamily Chandra\_repro} script performed the bad pixels removal and the instrument errors correction; afterwards, we removed the background flares and we used {\ttfamily Blanksky} background files, filtered and normalized to the count rate of the source hard X-ray image (9-12 keV), in order to subtract the background. The final exposure time is 7939 s.
\\ \indent
We identified and removed point sources using the CIAO tool {\ttfamily Wavdetect}. Comparing X-ray sources with optical counterparts, we checked the astrometry of the \textit{Chandra} data, and found it to be accurate; no registration correction was necessary. Unless otherwise stated, the reported errors are at 68$\%$ (1$\sigma$) confidence level.

\newpage
\subsection{Results}

\subsubsection{Surface Brightness Profile}
\label{sec:betamodel}

\begin{figure}[h]
	\centering
	\includegraphics[width=26em]{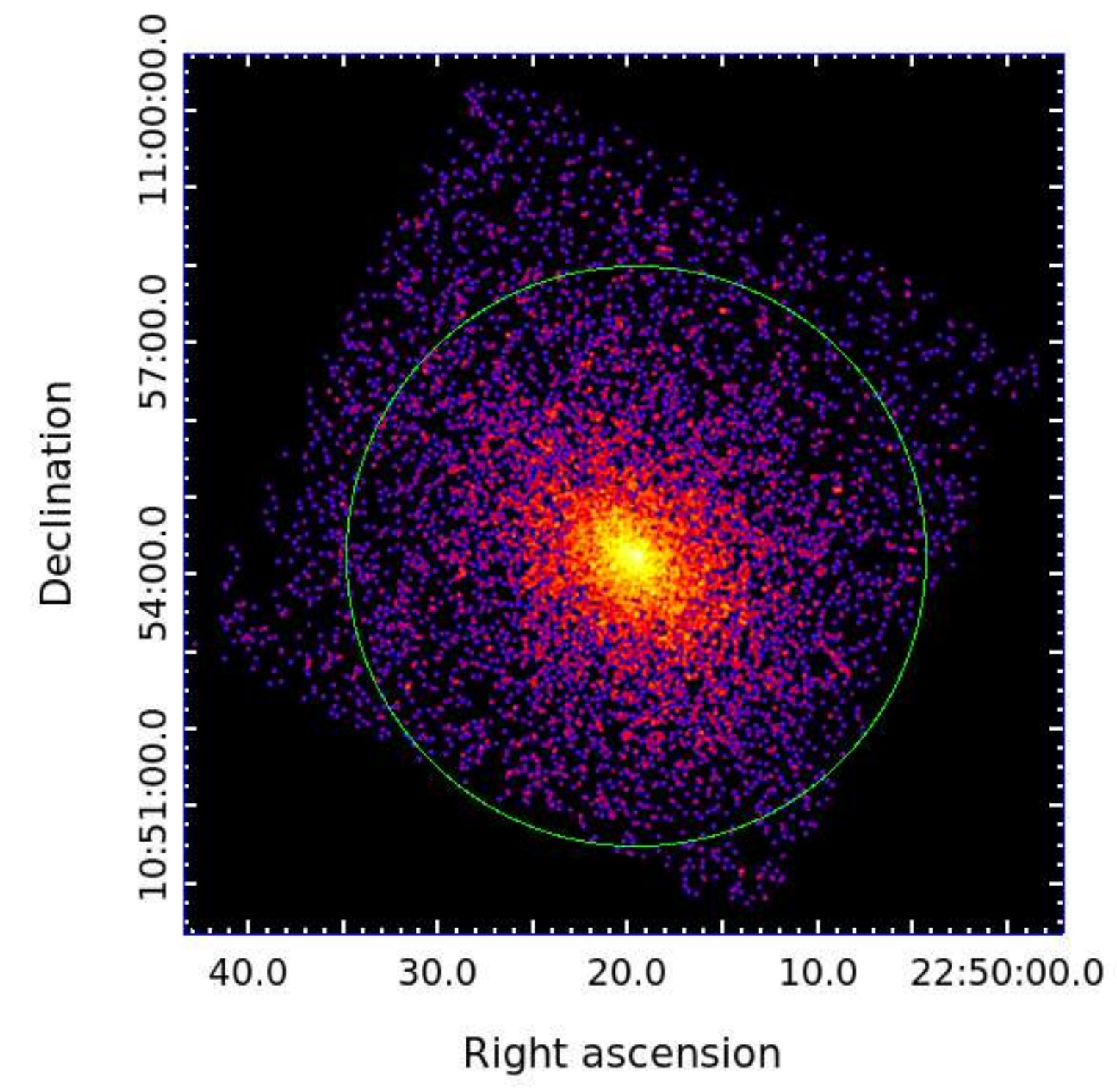}
	\caption{0.5-2 keV image of A2495, smoothed with a 5$\sigma$ gaussian. The green circle represents the maximum radius considered for the spectral analysis (see Sec. \ref{sec:xanalisis}).}
	\label{fig:A2495}
\end{figure}

\begin{figure}[h]
	\centering
	\includegraphics[height=23em, width=27em]{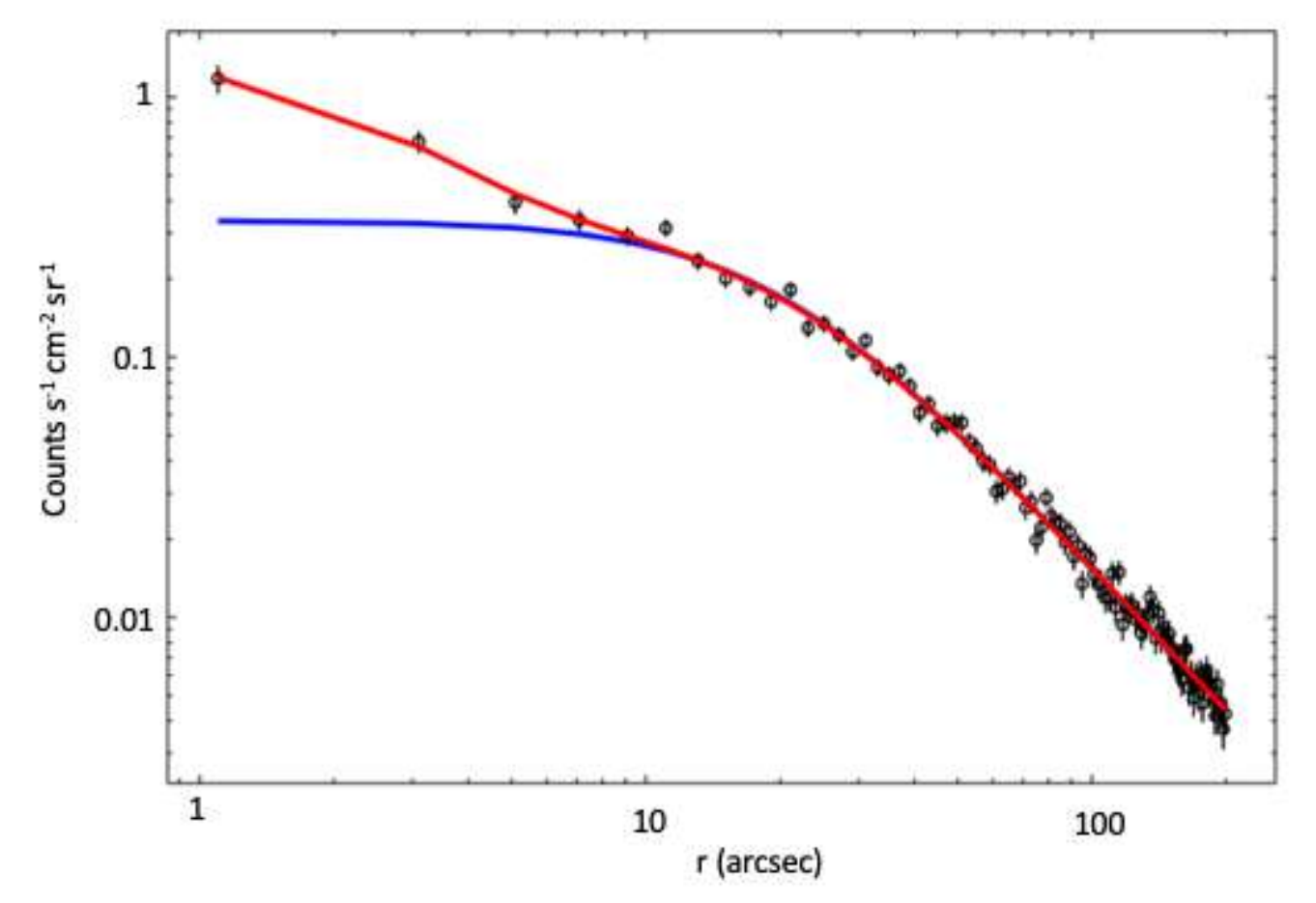}
	\caption{Surface brightness profile of A2495 fitted in the external regions ($\geq$ 30$''$) with a single $\beta$-model (blue line) and on every radius with a double $\beta$-model (red line). The blue line in the inner regions was obtained by extrapolating the model.}
	\label{fig:surbri}
\end{figure}

Figure \ref{fig:A2495} shows the smoothed 0.5-2 keV image of A2495. We produced a background-subtracted, exposure-corrected image and then extracted the surface brightness profile from a series of 2$''$-width concentric annuli centered on the X-ray peak.
\\ \indent
We used {\ttfamily Sherpa} in order to fit this profile in the outer radii ($\geq$ 30$''$) with a single $\beta$-model \citep[][]{Cavaliere-Fusco_1976}, and then extrapolated it to the inner regions. The best-fit values are {\ttfamily r0}=15.9 $\pm$ 0.9 arcsec, {\ttfamily beta}=0.46 $\pm$ 0.01 and {\ttfamily ampl}=0.41 $\pm$ 0.02 counts s$^{-1}$ cm$^{-2}$ sr$^{-1}$; the ratio between the $\chi^2$ and the degrees of freedom ($\textit{DoF}$) is $\chi^2/DoF \simeq$ 1.59. The result is represented with the blue line in Fig \ref{fig:surbri}.

We can clearly note the brightness excess characterizing the central regions of cool-core clusters. We therefore fitted a double $\beta$-model \citep[][]{Mohr_1999, LaRoque_2006} on the entire radial range. 
In this case, the best-fit parameters are {\ttfamily r0$_1$}=2.74
$^{+4.55}_{-1.63}$ arcsec, {\ttfamily beta$_1$}=0.64
$^{+1.61}_{-0.25}$ and {\ttfamily ampl$_1$}=1.09 $^{+0.55}_{-0.24}$
counts s$^{-1}$ cm$^{-2}$ sr$^{-1}$ for the first and {\ttfamily
  r0$_2$}=20.3 $^{+6.1}_{-2.1}$ arcsec, {\ttfamily beta$_2$}=0.48
$^{+0.06}_{-0.01}$ and {\ttfamily ampl$_2$}=0.31 $^{+0.04}_{-0.11}$
counts s$^{-1}$ cm$^{-2}$ sr$^{-1}$ for the second
$\beta$-Model. $\chi^2/DoF$ is 1.05. The corresponding model line is
shown in red in Fig. \ref{fig:surbri}. This suggests that  A2495 is, indeed, a
cool-core cluster, as expected from the selection criteria exploited for
the cluster choice.

\subsubsection{Spectral Analysis}
\label{sec:xanalisis}

 We extracted and fitted spectra of A2495 in the 0.5-7 keV band via
 {\ttfamily Xspec} (vv.12.9.1), excluding data above 7.0 keV and below
 0.5 keV in order to prevent, respectively, contamination from the
 background and calibration uncertainties. We derived the global
 cluster properties extracting a spectrum from a $\sim$ 200$''$
 circular region centered on the X-ray peak (located at
 $RA$=22$^{\text{h}}$50$^{\text{m}}$19.4$^{\text{s}}$,
 $DEC$=+10$\degree$54$^{\text{m}}$14.2$^{\text{s}}$). 
 The spectrum was then fitted assuming an absorbed emission produced by a collisionally-ionized diffuse gas, making use of the {\ttfamily wabs*apec} model. The hydrogen column density was fixed at $N_H$ = 4.73 $\cdot$ 10$^{20}$ cm$^{-2}$ \citep[estimated from][]{Kalberla_2005}; redshift was fixed at $z$ = 0.07923. The only parameters left free to vary were the abundance $Z$, the temperature $kT$ and the normalization parameter. We found $kT$ = 3.90 $\pm$ 0.20 keV, $Z$ = 0.54 $^{+0.11}_{-0.10}$ Z$_\odot$ and $F$(0.5-7 keV) = 1.07 $^{+0.01}_{-0.02} \cdot $ 10$^{-11}$ erg s$^{-1}$ cm$^{-2}$, leading to a total luminosity in the 0.5-7 keV band of $L$(0.5-7 keV) = (1.44 $\pm \ 0.02$)$ \ \cdot$ 10$^{44}$ erg s$^{-1}$. 
 \\ \indent
 We then performed a projected analysis, using a series of concentric rings covering the entire CCD and centered on the X-ray peak, each of which contains a minimum of 2500 total counts (the maximum radius reached, corresponding to 200$''$, is visible in Fig. \ref{fig:A2495}). Results are listed in Table \ref{tab:proje}.

\begin{table}[!htb]
	\scriptsize
	\centering 
	\begin{tabular}{c c c c c }
		\hline
		\hline
		$r_{\text{min}}$-$r_{\text{max}}$ &  Counts & kT & Z  \footnote{Solar abundance is estimated from the tables of \cite{Grevesse-Anders_1989}.} &  $\chi^2/Dof$
		 \\
		
		[arcsec] & & [keV] & [Z$_{\odot}$] &\\
		\hline
		1.5 - 25.1 & 2230 (98.9 \%) & 3.35 $^{+0.43} _{-0.29}$ & 1.04 $^{+0.43}_{-0.31}$ & 53/67 \\
		25.1 - 47.2 &  2240 (97.4 \%) & 4.05 $^{+0.55}_{-0.48}$ & 0.57 $^{+0.41}_{-0.28}$ & 94/67 \\
		47.2 - 72.8 &  2002 (94.1 \%) & 4.09 $^{+0.54}_{-0.47}$ & 0.46 $^{+0.29}_{-0.24}$ & 67/61 \\
		72.8 - 100.4 &  1718 (89.6 \%) & 3.97 $^{+0.65}_{-0.58}$ & 0.34 $^{+0.32}_{-0.23}$ & 53/55 \\
		100.4 - 129.9 &  1513 (82.1 \%) & 4.75 $^{+1.20}_{-0.89}$ & 0.36 $^{+0.42}_{-0.35}$ & 36/49 \\
		129.9 - 157.4 &  1368 (76.7 \%) & 5.30 $^{+1.34}_{-1.05}$ & 0.82 $^{+0.81}_{-0.55}$ & 45/45 \\
		157.4 - 185.5 &  1151 (67.1 \%) & 5.01 $^{+2.07}_{-1.26}$ & 0.44 $^{+0.88}_{-0.44}$ & 29/39 \\
		185.5 - 214.0 &  1020 (56.6 \%) & 3.09 $^{+1.65}_{-0.86}$ & 6.4e-02 & 41/35 \\
		\hline
	\end{tabular}
	\caption{Fit results of the projected radial analysis. The first column show the lower and upper limits of the extraction rings in arcsec, the second column represents the number of source photons coming from each ring, while the percentage indicates their number compared to the total photons of the same region. Finally, in the last three columns we report the values ​​of temperature, metallicity  and $\chi^2/$ DoF, respectively.} \label{tab:proje}
\end{table}

The low statistics led to large uncertainties in the measured abundances; therefore, in this work the cluster metallicity will not be discussed.

\begin{figure}[h]
	\centering
	\includegraphics[height=23em, width=26em]{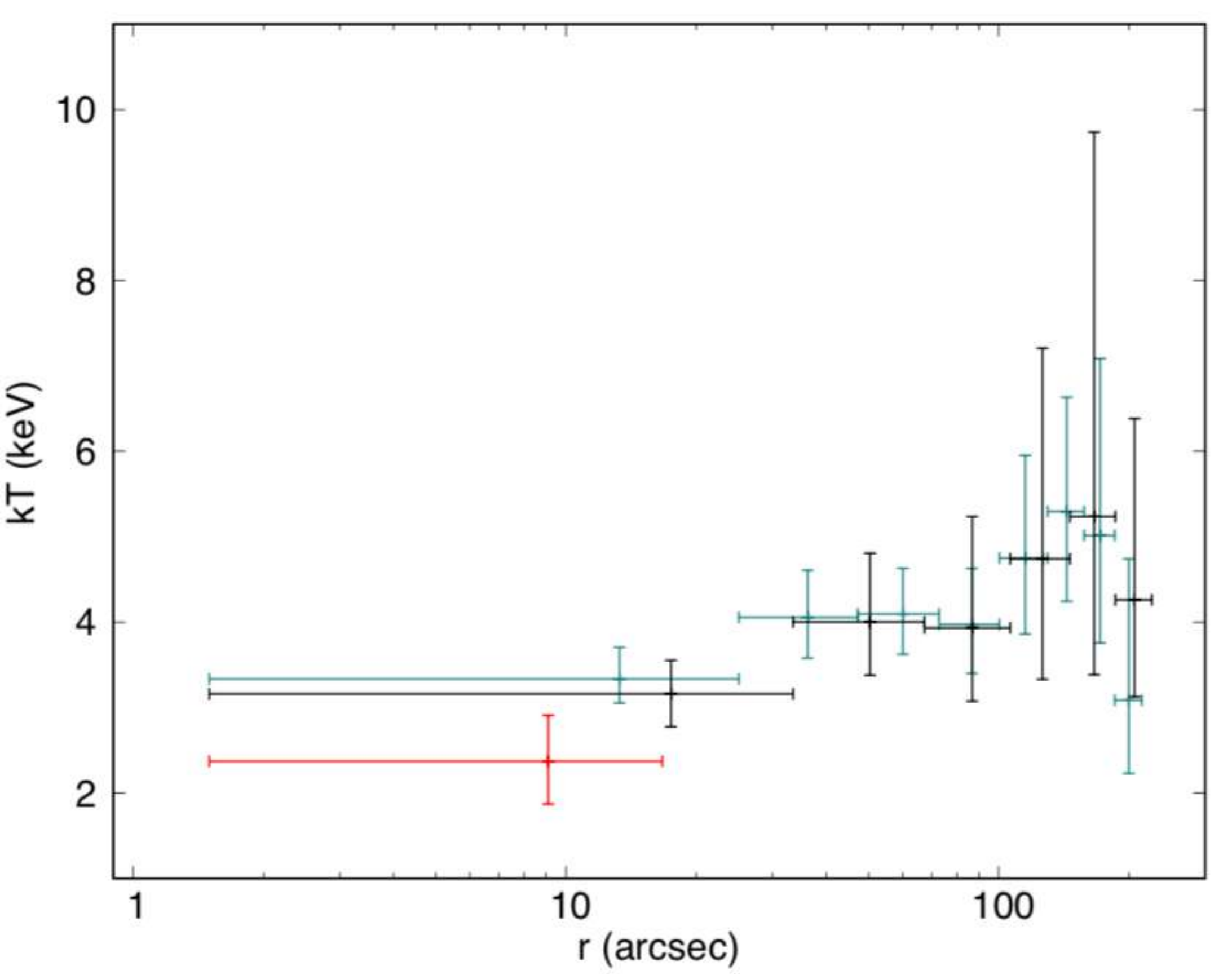}
	\caption{Projected (cyan) and deprojected (black) temperature profile of A2495. Bars in the x-axis represent the limits of the extraction rings, while the y-axis ones are the errors for the temperature values. The red point is derived from the spectral analysis described in Sec. \ref{sec:cavities}.}
	\label{fig:temp}
\end{figure}

The projected temperature profile of A2495 is shown in cyan in Fig \ref{fig:temp}. We carried out the same analysis making use of elliptical rings, but we did not find any significant difference.
\\ \indent
In order to account for the projection effects, we performed a deprojection analysis by adopting the {\ttfamily projct} model. For this purpose, we used concentric rings containing at least 3500 total counts. Similarly as above, the X-ray peak was excluded. Spectra were fitted using a {\ttfamily projct*wabs*apec} model, in which temperature, abundance and normalization parameters were left free to vary. Results are listed in Table \ref{tab:depro}. The deprojected temperature profile of the cluster is shown in black in Fig \ref{fig:temp}.

\begin{table*}[!htb]
	\normalsize
	\centering 
	\begin{tabular}{c c c c c c c c}
		\hline
		\hline
		$r_{min}$-$r_{max}$ & Counts & kT & Z  &  N(r) (10$^{-4}$) & Electronic Density & Pressure & t$_{cool}$   \\ 
		
		[arcsec]  & & [keV] & [Z$_{\odot}$] & & [$10^{-2} \ $cm$^{-3}$] & [10$^{-11}$ dy cm$^{-2}$] & [Gyr] \\
		\hline
		1.5 - 33.5  & 3143 (98.7 \%) & 3.16 $^{+0.39} _{-0.38}$ & 1.09 $^{+0.57}_{-0.38}$ & 14.3 $^{+2.1}_{-2.1}$ & 1.30 $^{+0.03}_{-0.03} $ & 11.8 $^{+1.7}_{-1.7} $ & 4.0 $^{+0.6}_{-0.6} $ \\
		33.5 - 67.4  & 2945 (95.7 \%) & 4.00 $^{+0.80}_{-0.62}$ & 0.43 $^{+0.39}_{-0.31}$ & 21.0 $^{+2.5}_{-2.5}$ & 0.57 $^{+0.03}_{-0.03} $ & 6.8 $^{+1.7}_{-1.4}$ & 10.0 $^{+2.5}_{-2.0} $\\
		67.4- 106.7 &  2430 (89 \%) & 3.92 $^{+1.30}_{-0.86}$ & 0.30 $^{+0.50}_{-0.30}$ & 17.5 $^{+2.8}_{-2.8}$ & 0.29 $^{+0.03}_{-0.03} $ & 3.3 $^{+1.4}_{-1.0}$ & 20.1 $^{+8.7}_{-6.4} $ \\
		106.7 - 146.1 & 2042 (80.2 \%) & 4.79 $^{+2.47}_{-1.41}$ & 0.32 $^{+0.85}_{-0.32}$ & 15.2 $^{+2.7}_{-3.3}$ & 0.18 $^{+0.03}_{-0.03} $ & 2.6 $^{+1.8}_{-1.2}$ & 34.0 $^{+22.6}_{-16.1} $ \\
		146.1 - 186.0 & 1709 (69.1 \%) & 5.19 $^{+4.50}_{-1.85}$ & 1.46 $^{+1.47}_{-1.47}$ & 8.7 $^{+3.7}_{-3.5}$& 0.11 $^{+0.04}_{-0.03} $ & 1.6 $^{+1.9}_{-1.1}$ & 61.1 $^{+74.3}_{-42.2} $  \\
		186.0 - 225.8 & 1384 (56.3 \%) & 4.27 $^{+2.12}_{-1.13}$ & 0.16 $^{+0.54}_{-0.16}$ & 18.1 $^{+2.4}_{-2.9}$ & 0.12 $^{+0.02}_{-0.03} $ & 1.6 $^{+1.2}_{-0.7}$ & 46.9 $^{+32.4}_{-23.6} $  \\
		\hline
	\end{tabular}
	\caption{Fit results of the deprojection analysis. The first two columns present, respectively, the limits of the annular regions and the number of source photons coming from each ring, while the percentage indicates their number compared to the total photons of the same region. The remaining columns report temperature, metallicity, normalization factor, electronic density, pressure and cooling time. The fit gives $\chi^2/$DoF = 394/437.}
	\label{tab:depro}
\end{table*}

In both profiles, the temperature value rises, as expected, moving from the center to the outskirts. The outermost bin of the projected analysis suggests the decline typical of relaxed clusters \citep[][]{Vikhlinin_2005}. This is less pronounced in the deprojected profile, where the poor statistics led to larger errors.
\\ \indent
The normalization factor $N$(r) of the {\ttfamily apec} model (see Table \ref{tab:depro}), if calculated through the deprojection analysis, allows us to obtain an estimate of the electronic density. It is defined as:

\begin{equation}
N(\text{r}) = \dfrac{10^{-14}}{4\pi[D_A(1+z)]^2}\int n_en_p dV
\end{equation}
\label{norm}

where $D_A$ is the angular distance of the source, calculated as $D_A=\dfrac{D_L}{(1+z)^2}$, $n_e$ represents the electronic density, $n_p$ the proton density and \textit{V} is the shell volume. For a collisionally-ionized plasma \citep[e.g.,][]{Gitti_2012}:

\begin{equation}
\int n_en_p dV \simeq 0.82 n_e^2 V
\end{equation}

The electronic density can therefore be estimated as:

\begin{equation}
n_e=\sqrt{10^{14} \bigg(\dfrac{4 \pi \cdot N(r) \cdot [D_A \cdot (1+z)^2]}{0.82 \cdot V}\bigg)}
\end{equation}

\begin{figure}[h]
	\centering
	\includegraphics[height=22em, width=26em]{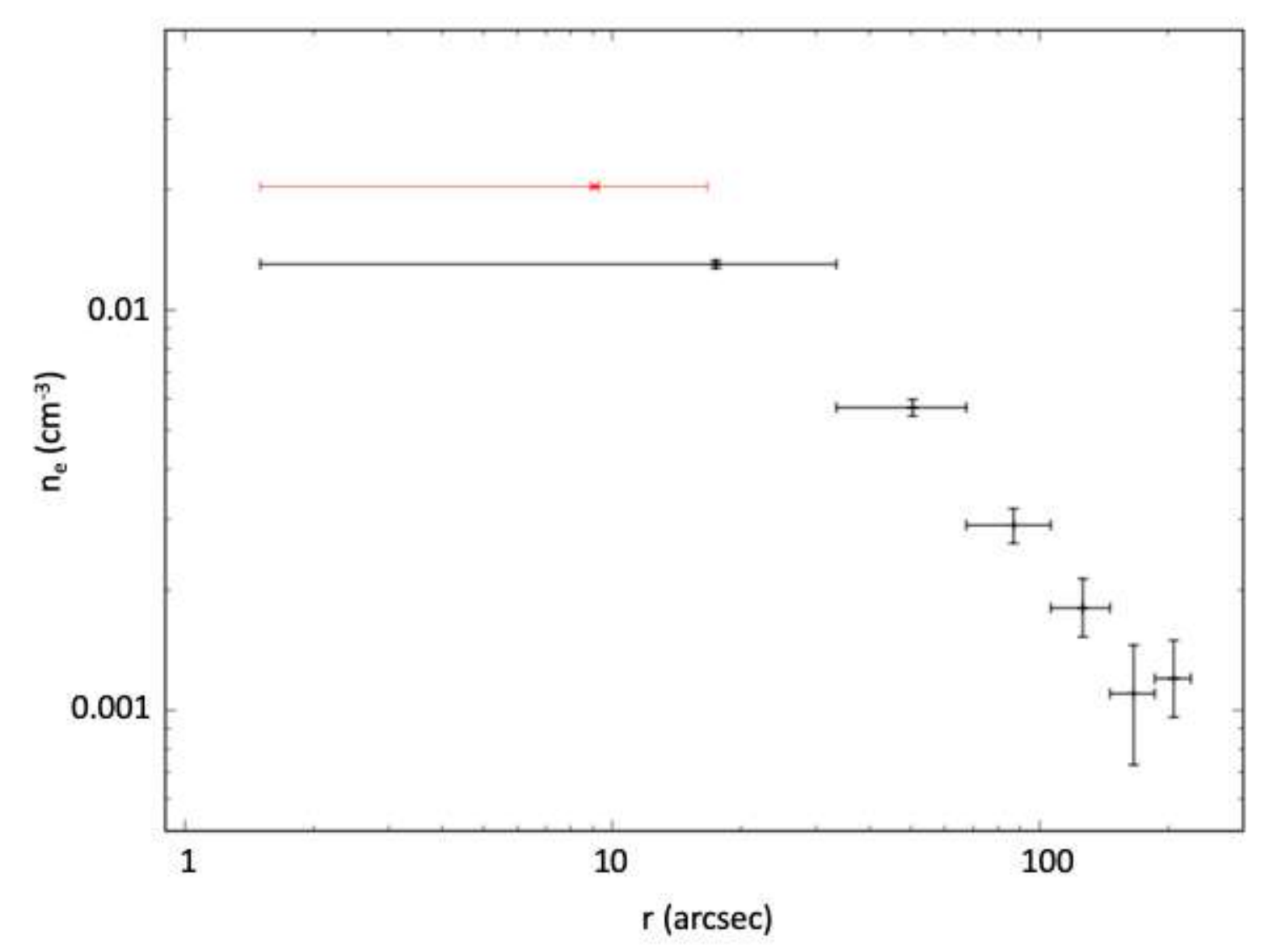}
	\caption{Density profile of A2495, obtained from the deprojection analysis. X-axis bars represent the limits of the extraction rings, while the y-axis ones are the errors estimated for the density. The red point is derived from the spectral analysis described in Sec. \ref{sec:cavities}.}
	\label{fig:density}
\end{figure}

Table \ref{tab:depro} lists the density values for each shell. The corrisponding radial density profile is shown in Fig \ref{fig:density}.

\begin{figure}[h]
	\centering
	\includegraphics[height=24em, width=25.3em]{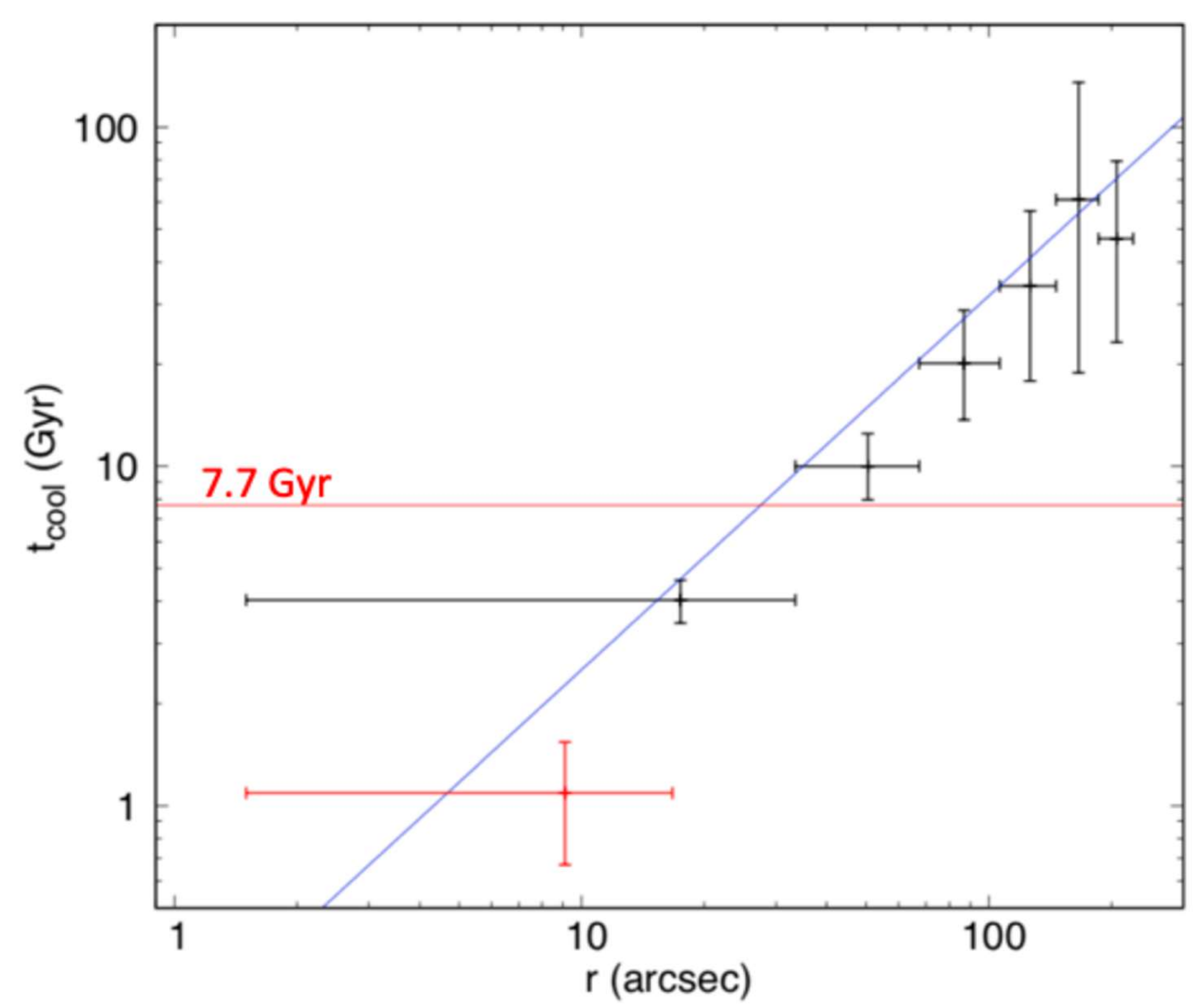}
	\caption{Cooling time profile for A2495. Each point represents the cooling time value for one of the annular regions used for the spectral extraction, and the x-axis errorbars are the lower and upper limits of each ring. The red line represents t$_{\text{age}}$ = 7.7 Gyr, while the blue line is the best-fit of the profile: t$_{\text{cool}}$=$0.199r^{1.102}$.  The red point is derived from the spectral analysis described in Sec. \ref{sec:cavities}.}
	\label{fig:coolt}
\end{figure}

From the temperature and density profiles we estimated the cooling time of each region, defined as:

\begin{equation}
t_{\text{cool}}=\dfrac{H}{\Lambda (T)n_e n_p}=\dfrac {\gamma}{\gamma-1}\dfrac{kT(r)}{\mu X n_e(r) \Lambda(T)}
\end{equation}
\label{eq:tcool}

where $\gamma$=5/3 is the adiabatic index, $H$ is the enthalpy, $\mu \simeq$ 0.61 is the molecular weight for a fully ionized plasma, $X$ $\simeq$ 0.71 is the hydrogen mass fraction and $\Lambda(T)$ is the cooling function \citep[][]{Sutherland-Dopita_1993}. Results are listed in Table \ref{tab:depro}, while the cooling time profile is shown in Fig. \ref{fig:coolt}.
\\ \indent
We adopted the definition of the cooling radius as the radius at which the cooling time is shorter than the age of the system. It is customary to assume the cluster's age to be equal to the lookback time at $z$=1, since at this time many clusters appear to be relaxed: $t_{\text{age}}$ $\simeq$ 7.7 Gyr. 

We estimated for A2495 a cooling radius of:

\begin{equation}
r_{\text{cool}} \simeq (28  \pm 16) \ \text{arcsec} \simeq (40 \pm 23) \ \text{kpc}
\end{equation}

In order to determine the X-ray luminosity produced within this radius ($L_{\text{cool}}$), we extracted a spectrum from a circular region with $r$ = $r_{\text{cool}}$ (excluding the central 1.5'') and fitted it with a {\ttfamily wabs*apec} model; Table \ref{tab:rcool} lists the results.

\begin{table}[!htb]
	\centering 
	\begin{tabular}{c c c  }
		\hline
		\hline
		$r_{\text{min}}$-$r_{\text{max}}$ & 1.5 - 28.5 & [arcsec]  \\
		Counts & 2500(98.9 \%) & \\
		kT & 3.29 $\pm ^{0.33}_{0.28}$ & [keV] \\
		Z & 0.86 $\pm ^{0.36}_{0.27}$ & [Z$_{\odot}$]  \\
		Flux$_{\text{Bolo}}$ & 3.17 & [10$^{-12}$ erg s$^{-1}$ cm$^{-2}$] \\
		Flux$_{2-10 \ \text{keV}}$ & 1.21 & [10$^{-12}$ erg s$^{-1}$ cm$^{-2}$]  \\
		\hline
	\end{tabular}
	\caption{Fit results to the cooling region. The $\chi^2$/Dof is 68/74.} \label{tab:rcool}
\end{table}

Therefore, the bolometric cooling luminosity is:

\begin{equation}
L_{\text{cool}}= 4.3 \ ^{+3.1}_{-3.4} \cdot 10^{43} \ \text{erg \ s}^{-1}
\end{equation}

The same spectrum was fitted with a {\ttfamily wabs*(apec+mkcflow)} model; {\ttfamily mkcflow} adds an isobaric multi-phase component, allowing us to reproduce a cooling flow-like emission, while {\ttfamily apec} takes into account the background contribution. Temperature and abundance of the {\ttfamily apec} model were left free to vary; we bounded them to the {\ttfamily high temperature} and {\ttfamily abundance} parameter of {\ttfamily mkcflow}. Redshift and absorbing column density were fixed at the Galactic values (see above), while the {\ttfamily low temperature} parameter of {\ttfamily mkcflow} was fixed at $\sim$ 0.1 keV; in this way, we are assuming a standard cooling flow. The {\ttfamily norm} parameter of {\ttfamily mkcflow} provides an estimate of the Mass Deposition Rate of the cooling flow ($\chi^2$/DoF $\sim$ 68/73): 

\begin{equation}
\dot{M} < 7 \ \text{M$_{\odot}$  yr$^{-1}$}
\end{equation}

An alternative method to obtain an estimate of the mass accretion rate exploits the luminosity $L_{\text{cool}}$ associated with the cooling region, assuming that it is all due to the radiation of the total gas thermal energy plus the \textit{pdV} work done on the gas as it enters the cooling radius. These assumptions are typical of the classical cooling flow model \citep{Fabian_1994}, that does not take into consideration heating produced by the central AGN. 

\begin{equation}
{\dot{M}} \simeq \dfrac{2}{5} \dfrac{\mu m_p}{kT} \cdot L_{\text{cool}} \simeq 52 \ ^{+38}_{-42} \ \text{M$_{\odot}$ yr$^{-1}$}
\end{equation}

\begin{figure}[h!]
	\centering
	\includegraphics[height=25em, width=26em]{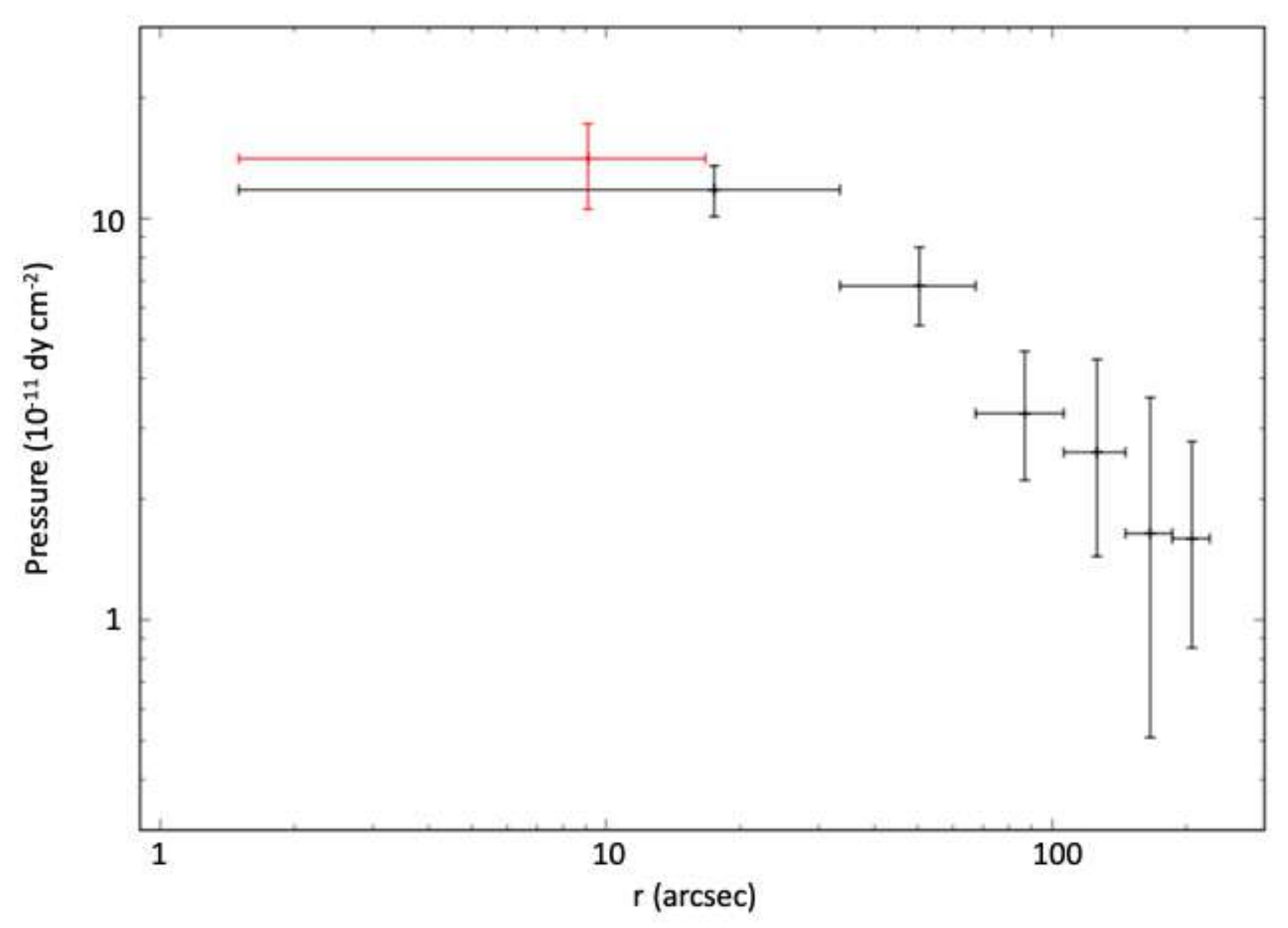}
	\caption{Pressure radial profile obtained from the deprojection analysis. The red point is derived from the spectral analysis described in Sec. \ref{sec:cavities}.}
	\label{fig:pressure}
\end{figure}

Finally, in Table \ref{tab:depro} we list the pressure values calculated as $p = 1.83n_e kT$, where $n_e$ was taken from the deprojection analysis. Fig. \ref{fig:pressure} shows the correspondent radial profile.

\section{Discussion}

\begin{figure*}[h]
	\hspace*{-6.9cm} 
	\centering
	\includegraphics[height=50em, width=96em]{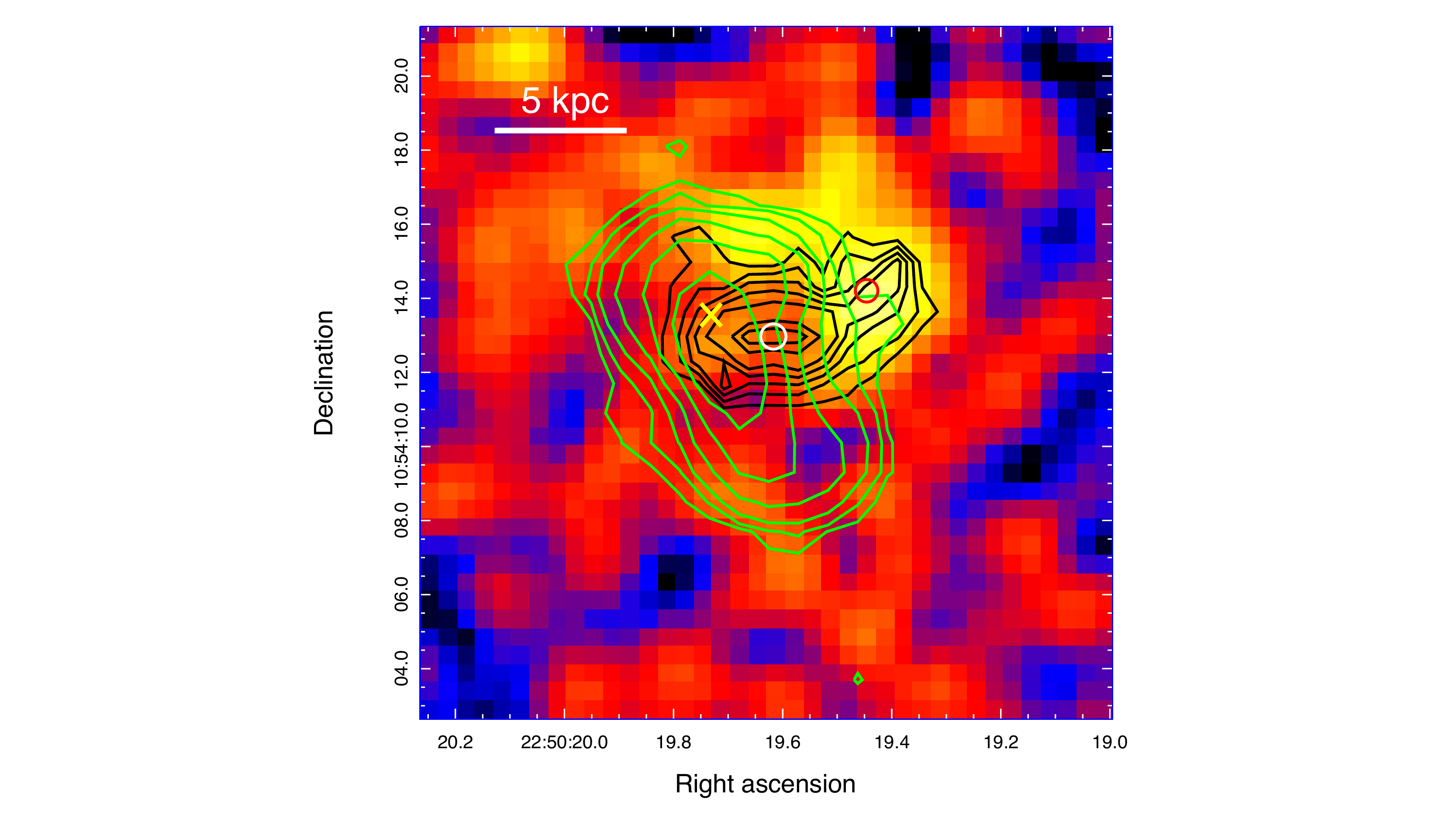}
	\hspace*{1cm}
	\includegraphics[height=20em, width=47em]{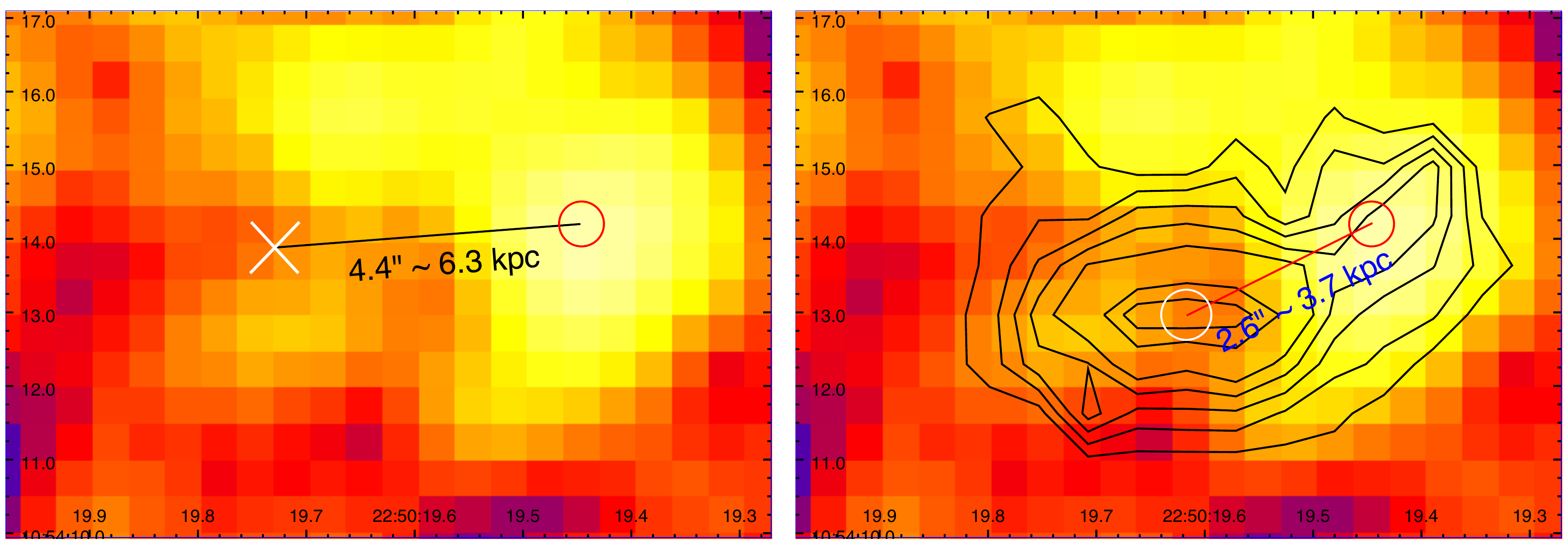}
	\caption{\textit{Top Panel}: 1.4 GHz (green) and H$\alpha$ (black) contours overlaid on the 0.5-2 keV \textit{Chandra} image of A2495. The red and white circles represent the X-ray and H$\alpha$ peaks, respetively, while the yellow cross is the position of the X-ray centroid. \textit{Bottom-Left Panel}: Offset between the emission centroid (white cross) and the X-ray peak (red circle).  \textit{Bottom-Right Panel}: Offset between the H$\alpha$ (white circle) and the X-ray (red circle) peaks.}
	\label{fig:multiwav}
\end{figure*}

In Fig \ref{fig:multiwav} we show the smoothed 0.5-2 keV image, with overlaid the 1.4 GHz contours, zoomed into the cluster center.

We highlight the presence of an offset between the BCG
($RA$=22$^{\text{h}}$50$^{\text{m}}$19.7$^{\text{s}}$,
$DEC$=+10$\degree$54$^{\text{m}}$12.7$^{\text{s}}$) and the X-ray peak
(red circle); this suggests that there is a relative motion between
them. By means of the X-ray isophotes, we estimated the centroid (bottom-left panel of Fig. \ref{fig:multiwav}, black circle), located at 
$RA$=22$^{\text{h}}$50$^{\text{m}}$19.7$^{\text{s}}$,  $DEC$=+10$\degree$54$^{\text{m}}$13.8$^{\text{s}}$. The determination was made calculating the centroid in several regions centered on the peak and with radii ranging from 50 to 200 kpc.

The offset approximately measures 4.4$''$ $\pm$ 1.0$''$, corresponding to 6.3 $\pm$ 1.4 kpc. We investigated the possibility of the presence of a point source in the position of the X-ray peak, that could bias its location, extracting a spectrum from a $\sim$ 10$''$ radius region centered on the peak. The radius was chosen in order to obtain enough statistics to produce a reliable fit. The spectrum was then fitted with two models: {\ttfamily wabs*apec} and {\ttfamily wabs*(apec+powerlaw)}. The {\ttfamily powerlaw} component, added to model the point-source emission, has two parameters: the spectral index and a normalization factor. If a point source is present, we expect a significant improvement of the fit using the second model. The first produced $\chi^2/DoF$=42/54, while for the second $\chi^2/DoF$= 40/52. In order to determine if the distribution improved with the addition of the {\ttfamily powerlaw} emission, we applied the F-stat method, obtaining $P_f$=1.3\footnote{$p$ = 0.28, corresponding to a null hypothesis probability of P = 1-$p$ = 0.72.}, thus indicating that the addition of a second component is not statistically significant. 
We then performed a search by coordinates in X-ray, optical and infrared catalogues. However, the closest object seems to be an IR source (SSTSL2 J225019.00+105414.5) located more than 10 kpc ($\sim$ 7.5$''$) away from the peak; therefore, it can not represent its optical counterpart. We thus find no evidence of an X-ray point source at the position of the X-ray peak, and conclude that the offset is real.
\\ \indent
We propose that this offset could be produced by \textit{sloshing}, an oscillation of the ICM within the cluster potential well, generated from perturbations such as, for example, minor mergers. This mechanism is usually tied to the formation of \textit{cold fronts} \citep[][]{Markevitch-Vikhlinin_2007}, discontinuities between the hot cluster atmosphere and a cooler region (i.e. the cooling center) that are moving with respect to each other. In this scenario, the oscillation of the ICM has displaced the X-ray peak from the cluster's centre: the cooling process is not taking place on the BCG nucleus (see Sec. \ref{sec:intro}). However, deeper observations are necessary to test this hypothesis by means of detailed analysis of the thermodynamic properties of the galaxy cluster central regions.

\subsection{H$\alpha$ emission analysis}
\label{sec:halpha}

\begin{figure}[h]
	\centering
	\includegraphics[width=26.5em]{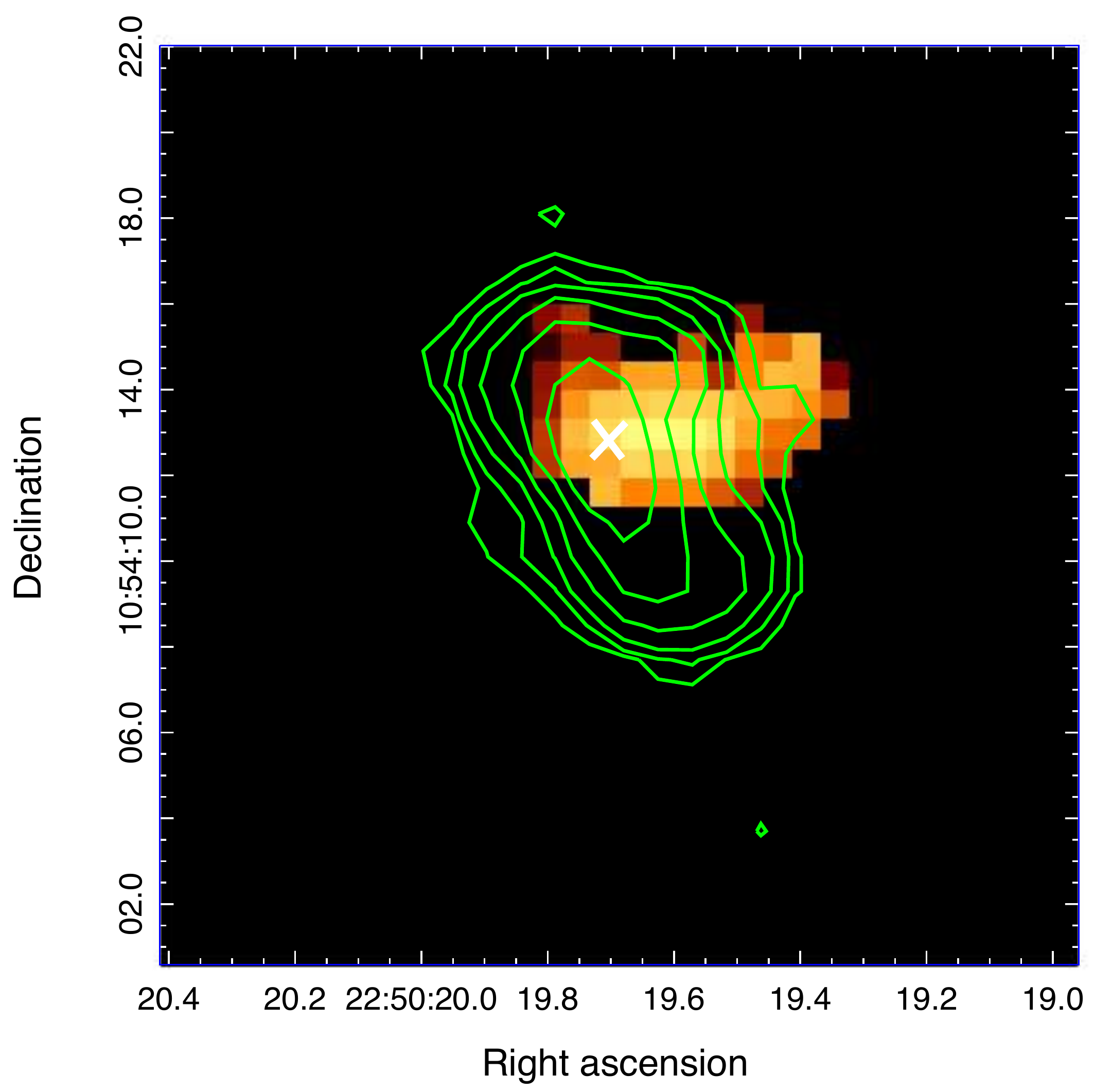}
	\caption{H$\alpha$ emission structure in the inner regions of A2495 observed by \textit{VIMOS}  \citep[][]{Hamer_2016}. The green radio contours are the same of Fig. \ref{fig:multiwav} and \added{the white cross represents the BCG center, located in $RA$ = 22$^\text{h}$50$^\text{m}$19.7$^\text{s}$, $DEC$ = +10$\degree$54$^\text{m}$12.7$^\text{s}$ (J2000). The mean seeing of the H$\alpha$ image is 0.95$''$, and its} units are in 10$^{-16}$ erg s$^{-1}$ cm$^{-2}$ \AA$^{-1}$.}
	\label{fig:halpha}
\end{figure}

The detection of line emitting nebulae in the proximity of BCGs is known to be an indicator of the presence of multiphase gas. Such structures are only found when the central entropy drops below 30 keV cm$^2$ or, equivalently, when t$_{\text{cool}} \ <$ 5 $\cdot$ 10$^8$ yr \citep[e.g.,][]{Cavagnolo_2008, McNamara_2016}.
\textit{VIMOS} (Visibile Imaging Multi-Object Spectrograph) observations of the H$\alpha$ line emission of a sample of 73 clusters, including A2495, were presented by \citet{Hamer_2016}. To be consistent with the multi-wavelength data shown in this work, we checked and corrected the astrometry exploiting the HST F606W image (see Sec. \ref{sec:optical}) and the F555W image that was used by \citet{Hamer_2016} to align the VIMOS data, finding an offset of ~1.3 arcsec between the BCG centroids in them. In Fig. \ref{fig:halpha} we show the astrometrically corrected H$\alpha$ image.

The total luminosity is (5.03 $\pm$ 0.81) $\cdot$ 10$^{39}$ erg s$^{-1}$. The structure is elongated on a scale of $\sim$ 5.6$''$ ($\sim$ 8 kpc). In Fig. \ref{fig:multiwav} we show the H$\alpha$ emission contours (black) overlaid on the smoothed 0.5-2 keV image. 
The H$\alpha$ structure stretches towards and surrounds the X-ray peak, connecting it to the BCG. The same behaviour is found in other galaxy clusters (e.g., A1991, A3444 and RXJ0820.9+0752 in \citealt{Hamer_2012, Bayer-kim_2002}, or A1111, A133, A2415 and others in \citealt{Hamer_2016}), where the H$\alpha$ plume either follows the X-ray peak or acts like a bridge between the peak and the BCG. This could suggest that the line emission is not produced by ionisation coming from stellar radiation or from the AGN; in this scenario, in fact, one expects the emission to be associated with the BCG, rather than with the X-ray core \citep[e.g.,][]{Hamer_2016, Crawford_2005}. This 10$^4$ K gas is therefore likely connected to the cooling ICM  \citep[e.g.,][]{Bayer-kim_2002, Canning_2012, Crawford_2005, Ferland_2009}. Moreover, we note that the A2495 line emission luminosity is significantly weaker (typically of about an order of magnitude) with respect to the plumed population presented in \citet{Hamer_2016}; this cluster could represent a lower luminosity limit of this sample.
\\ \indent
We also highlight the presence of a further, less pronounced ($\sim$
2.6$''$, corresponding to 3.7 kpc) offset between the H$\alpha$ and
the X-ray peaks. We first checked if it could be produced by
astrometric uncertainties. The VIMOS data were obtained with a seeing
of 0.95$''$ \citep[from][]{Hamer_2016}, while the overall 90\% uncertainty circle of {\it Chandra} X-ray absolute position has a
radius of 0.8$''$. The upper limit of the astrometric uncertainty is thus expected to be 
$\Delta x_{tot}=\sqrt{\Delta x_{Chandra}^2+\Delta x_{VIMOS}^2} =
1.24'' \simeq 1.7 \ \text{kpc}$; since our offset is larger ($\sim$
3.6 kpc), it is likely to be a real feature. 
\\ \indent
We argue that this offset could originate to the different
hydrodynamic conditions of the ICM: the H$\alpha$ emission is
generated from $\sim 10^{4}$ K gas, $\sim 10^2 - 10^3$ denser than
the surrounding hot ICM. Thus, hydrodynamic processes may lead to
differential motion between these two phases and generate the offset.
Another hypothesis is that thermal instabilities (or so-called {\it
 precipitation}) occurring in the inner regions of the cluster could
have produced the $10^{4}$ K gas \textit{in situ}, following the
chaotic cold accretion scenario (CCA) \citep[e.g.,][see
Sec. \ref{sec:cavities}]{Gaspari_2012, Voit_2015}.
\\ \indent
In some cases \citep[e.g., A1991,][]{Hamer_2012}, two H$\alpha$ peaks, with one being coincident with the X-ray one, are detected. As this object does not show a second, or even dominant, peak of line emission at the X-ray peak, it is possible this represents an earlier stage of these offsets, one in which there has not yet been sufficient time for the cooled gas at the offset location to grow to a mass comparable to that already in the BCG. The plume of optical line emission shows a very clear velocity gradient (see \citealt{Hamer_2016} for further details), and as the ionised gas can be used as a direct tracer of the motion of the cold gas, can be used to study the dynamics of the cold gas during the sloshing process.  Following the method of \citet{Hamer_2012}, we estimated the dynamical offset timescale of the cold gas.  The plume extends for $\sim$ 8 kpc from the centre of the BCG (5.6 arcsec at 1.44 kpc/$"$), the gas velocity at the centre of the BCG is consistent with the stellar component to within 10 km $\cdot$ s$^{-1}$, indicating that the plume really extends out from the BCG and is not just a projection effect.  The velocity of the gas changes smoothly and consistently along the plume, and we measured a velocity difference of $\sim$ +350 km between the gas at the end of the plume and at the centre of the BCG. Thus, we measured a projected extent D' = 8 kpc = 2.47 $\cdot$ 10$^{17}$ km and a projected velocity shift of V' = +350 km $\cdot$ s$^{-1}$, indicating a projected timescale of T' = D'/V' = 7.05 $\cdot$ 10$^{14}$ s or $\sim$ 22.4 Myr.  To correct for the projection effects we must know the inclination of the offset (as T = T' $\times$ cos[i]/sin[i]).  While the inclination cannot be determined from the data in hand, the most likely inclination to expect is $\sim$ 60$\degree$ \citep[][]{Hamer_2012}, which would give an offset dynamical timescale of T$\sim$ 13 Myr. 

\subsection{Optical analysis}
\label{sec:optical}

\begin{figure*}[ht]
	\centering
	\includegraphics[height=26em, width=26em]{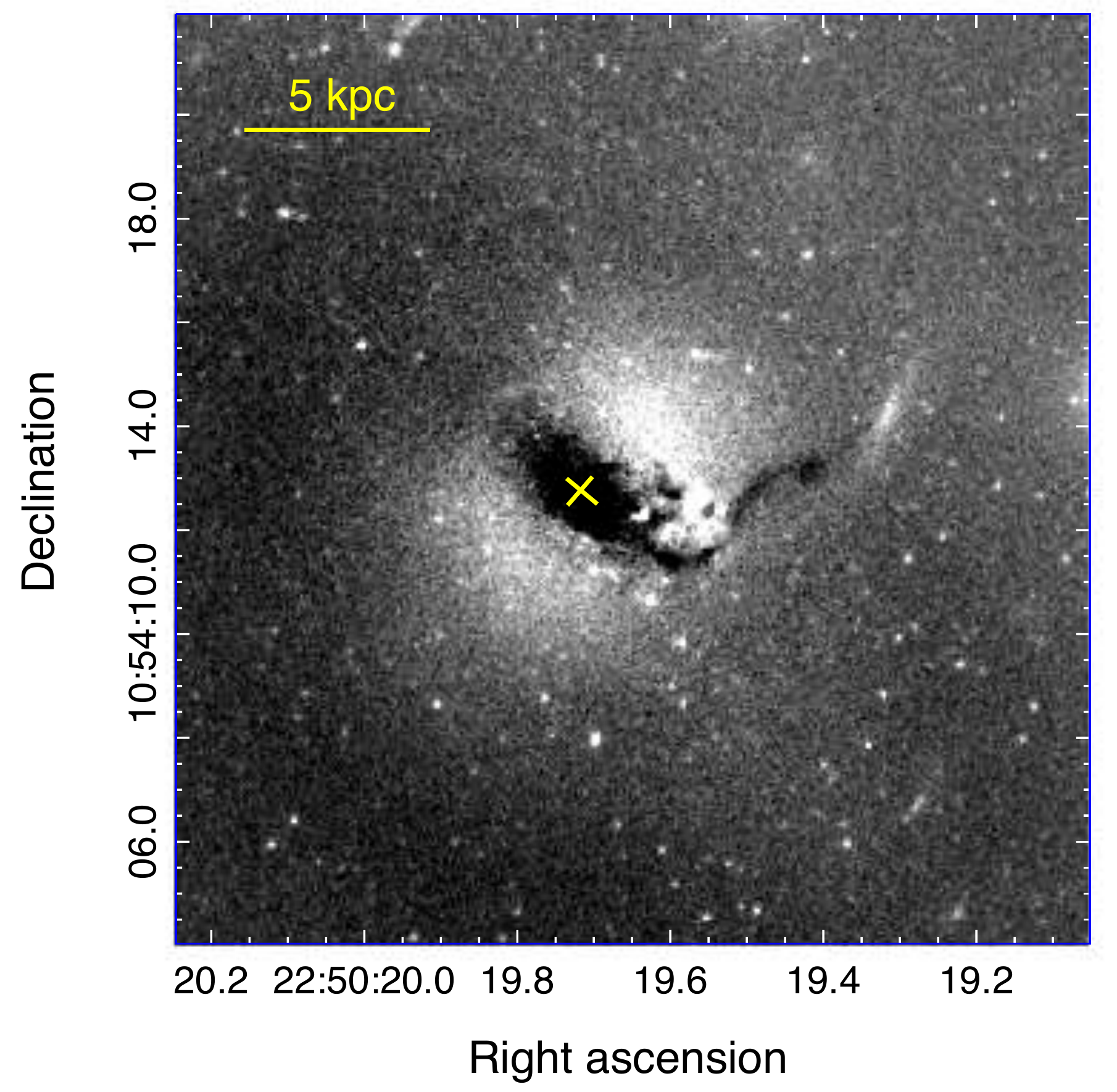}
	\includegraphics[height=26em, width=26em]{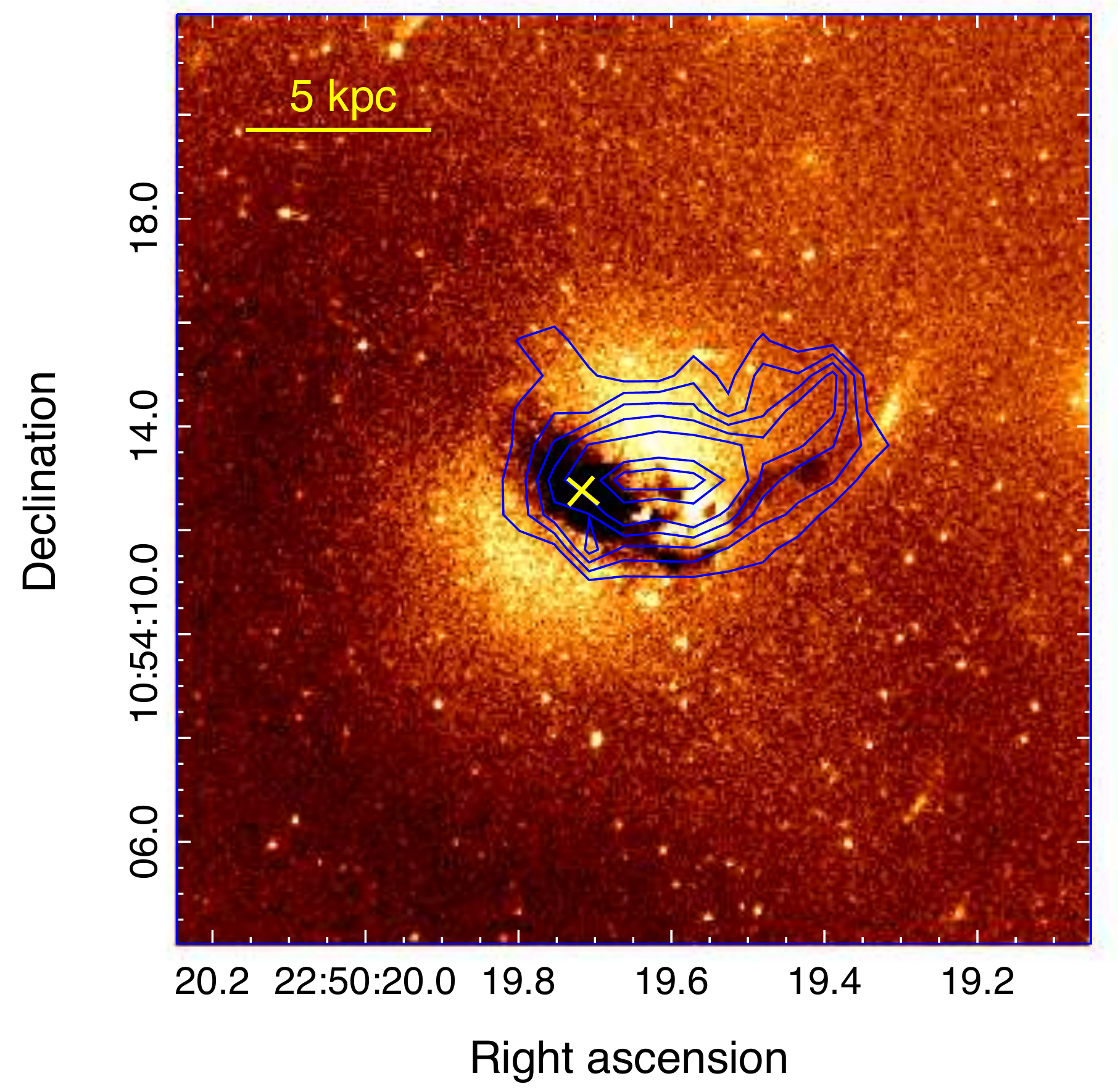}
	\caption{\textit{Left panel}: Dust extinction map of the BCG and core of A2495, obtained from the F606W HST image. Dust in absorption, seen as a filamentary structure, encircles the inner region of the galaxy and stretches towards the outskirts. \added{The yellow cross represents the BCG optical center.} \textit{Right panel}: H$\alpha$ contours overlaid on the dust extinction map. The two structures are co-spatial.}
	\label{fig:vband}
\end{figure*}

Deep optical images of the cluster galaxy were retrieved from the HST archive. We used HST ACS observations taken in the two wide filters F606W (V-band) and F814W (I-band). Images in these two filters have been instrumental in generating a dust extinction map and in the evaluation of the galaxy luminosity within the central 10kpc. In order to estimate the dust mass in the central region, we developed a 2-D galaxy model by fitting the starlight with elliptical isophotes in an iterative process with subsequent removal of sources in the field or by masking features not related to the stellar light distribution.  An extinction map $ A_V = -2.5~log(I_{obs} / I_{model})$  (Fig. \ref{fig:vband}) was generated from the observed intensity $(I_{obs})$ and the intensity of the starlight model $(I_{model})$ \footnote{We also tried an alternative method, performing a ratio of the F606W and F814 images, in order to avoid the dustiest regions in the galaxy, that could affect the starlight fitting process. However, we only found a difference of $\sim$ 20\% between the results of the two methods.}. To avoid the introduction of artifacts in the final map, particular care has been taken in the generation of the starlight model in the central few arc seconds, where a dust lane crosses the galactic center, making the isophote fitting process quite challenging. 

\begin{figure*}[ht]
	\centering
	\includegraphics[height=28em, width=55em]{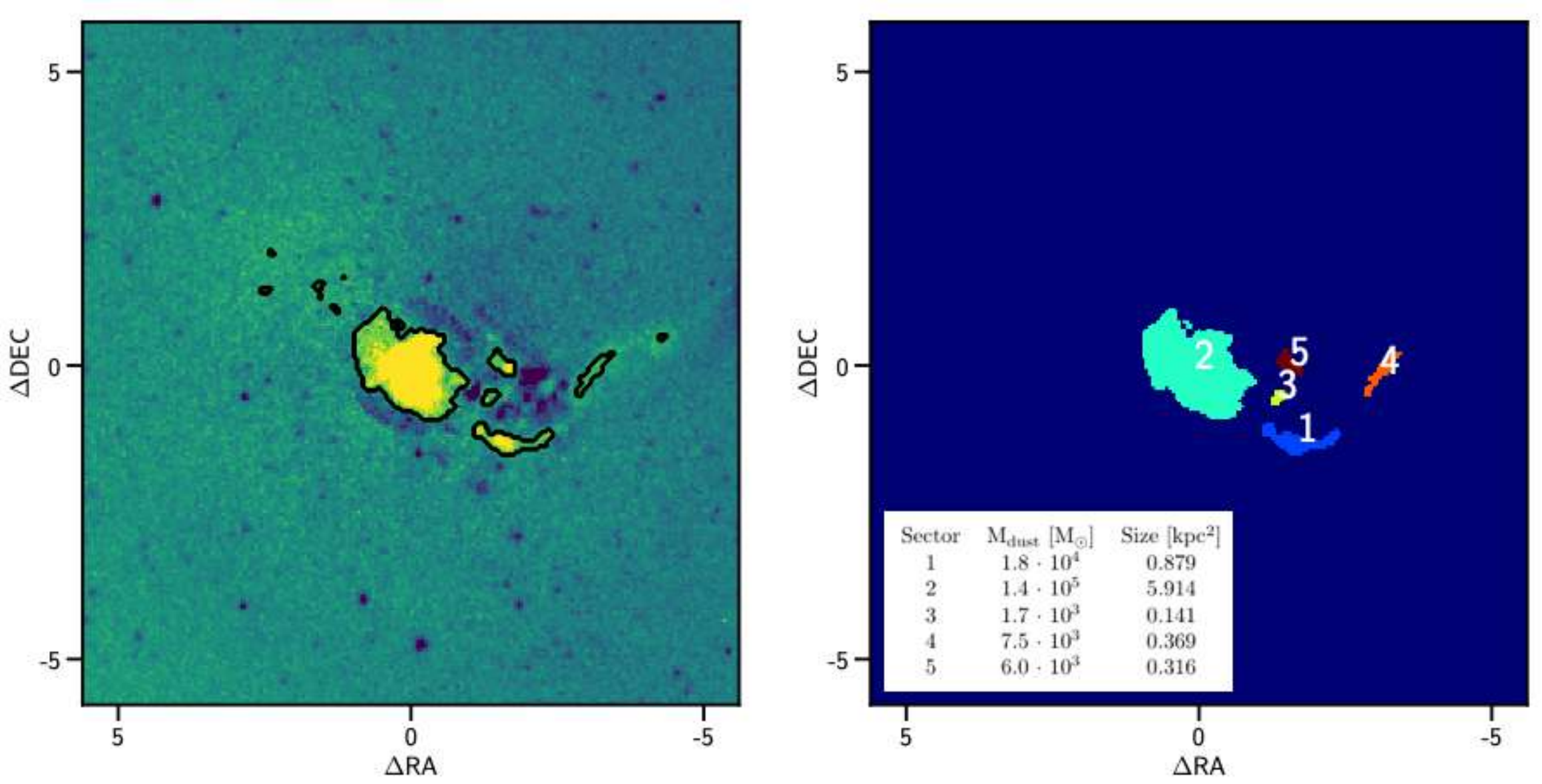}
	\caption{Sectors in which the dust mass estimation was performed. The table shows the dust mass values and the size for each sector. \added{$\Delta RA$=0, $\Delta DEC$=0 corresponds to the BCG optical center (see Sec. \ref{sec:radio})}.}
	\label{fig:sectors}
\end{figure*}

The extinction map confirms that an highly absorbed dust lane crosses the galaxy center, while a large wavy filamentary structure extends to the West for about $4^{\prime \prime}$. Additional knots of highly absorbed dust are apparent in the central kpc of the galaxy. Although the peak of the warm ionized gas does not correspond to the galactic center or any other location of high extinction, the lower ridge of the H$\alpha$ emission seems to be associated with dust absorption revealed in the filamentary structure. 
\\ \indent
Dust structures are common in BCGs. \citet{VanDokkum_1995} detected dust in 50\% of the galaxies included in their sample of 64 objects; \citet{Laine_2003} found signs of dust absorption in 38\% of 81 BCGs, classifying them, according to their morphology, into nuclear dust disk, filaments, patchy dust, dust rings and dust spirals (see Section 3.2 and Fig. 3 of \citealt{Laine_2003} for further details). The structure in A2495 seems to fall into the second class, looking like a classic dust filament. The BCG of this cluster was also part of the large Spitzer
dust survey presented by \citet{Quillen_2008} and discussed in \citet{O'Dea_2008}; these studies highlighted that dust in A2495 does not show a particularly remarkable mid-IR continuum and is not detected at 70 $\mu$m, likely suggesting that the star formation does not significantly heat the dust as it is, instead, seen in other systems (see \citealt{Quillen_2008} for more details).
\\ \indent
From the extinction map we calculated the total dust mass in the central 7 kpc under some basic assumptions for the grain size distribution and composition.  The approach and relative calculations are outlined in \citet{Goudfrooij_1994}; here we assume dust extinction properties similar to what is observed in the Milky Way. We used a grain size distribution proportional to a$^{-3.5}$ (where a is the grain radius) \citep{Mathis_1977, Goudfrooij_1994} and a mixture of graphite and silicate (equal absorption from those) with a specific grain mass density of 3 g/cm$^3$. Lower and upper limits for the grain size distribution were set to $a_- = 0.005~\mu$m and $a_+ = 0.22 ~\mu$m \citep{Draine_1984}.
The filamentary structure was divided into 5 sectors, as shown in Fig. \ref{fig:sectors}.

The total dust mass accounted in all the filamentary structures in the
central 7  kpc is $1.7 \cdot 10^5$ M$_\sun$. The reported mass should
be regarded as lower limit, considering that the method used in
generating the extinction map would mask a centrally symmetric diffuse
dust component, if present. Given the uncertainties in the generation
of an accurate galaxy model at the galactic center and the statistical
error, we estimate a 30\% uncertainty in the reported dust mass. We
remark that the optically detected dust likely underestimates the
total dust content \citep[see for example][for a study of
dust in a sample of elliptical galaxies]{Temi_2004}. 
\\ \indent
Assuming a dust/gas ratio of $M_{\text{dust}}/M_{\text{gas}}
\sim 1/100$ \citep[e.g.,][]{Edge_2010}, we expect for A2495 a minimum
cold gas mass within the central 7 kpc of $\sim$ 2 $\cdot$ 10$^7$
M$_\sun$ (again, with a 30\% uncertainty).  We note that the mass of the
H$\alpha$-emitting gas within the same region (assuming that it is
optically thin and in pressure equilibrium with the local ICM) is:

\begin{equation}
M_{\text{H$\alpha$}} \simeq L_{\text{H$\alpha$}} \dfrac{\mu m_p}{n_{\text{H$\alpha$}}\epsilon_{\text{H$\alpha$}}}
\end{equation}

where $L_{\text{H}\alpha}$ is that reported in Sec. \ref{sec:halpha}
and $\epsilon_{\text{H}\alpha}  \sim 3.3 \cdot 10^{-25}$ erg cm$^3$
s$^{-1}$ is the H$\alpha$ line emissivity; $n_{\text{H$\alpha$}}$ is
obtained assuming:

\begin{equation}
n_{\text{H$\alpha$}}T_{\text{H$\alpha$}} \simeq n_{\text{ICM}}T_{\text{ICM}}
\end{equation}

where $T_{\text{H$\alpha$}} \sim 10^4$ K, $T_{\text{ICM}}$ is the first value reported in Tab. \ref{tab:1500} (since the H$\alpha$ structure is located within the central 7 kpc), while $n_{\text{ICM}} \sim 1.83n_e$. Again, $n_e$ can be retrieved from Tab. \ref{tab:1500}. We thus obtain $M_{\text{H$\alpha$}} \sim 10^5 \pm 10^4$ M$_\odot$.
\\ \indent
The H$\alpha$ mass can account for just a negligible fraction of the
total cold gas mass estimated above: a significant amount is still
missing. We speculate that colder, likely molecular gas is present in
the central regions of A2495. This is supported by the correlation
between the H$\alpha$ luminosity and the molecular mass
\citep{Edge_2001, Pulido_2018}, from which we obtain $M_{\text{mol}} \sim$ 10$^9$ M$_\sun$. The contrast with the estimate determined from the dust/gas ratio is only apparent: the latter takes into account the dust mass within the central 7 kpc of the BCG and constitutes, per se, a lower limit (see above). Cold molecular gas could also lie offset from the BCG, as observed in other systems where evidence for CO line emission coincident with the offset X-ray peak was detected \citep[e.g.,][]{Hamer_2012}.

\subsection{Does sloshing regulate the generation of multiple cavity systems?}
\label{sec:cavities}

\begin{figure}[h]
	\centering
	\includegraphics[height=22em, width=26.5em]{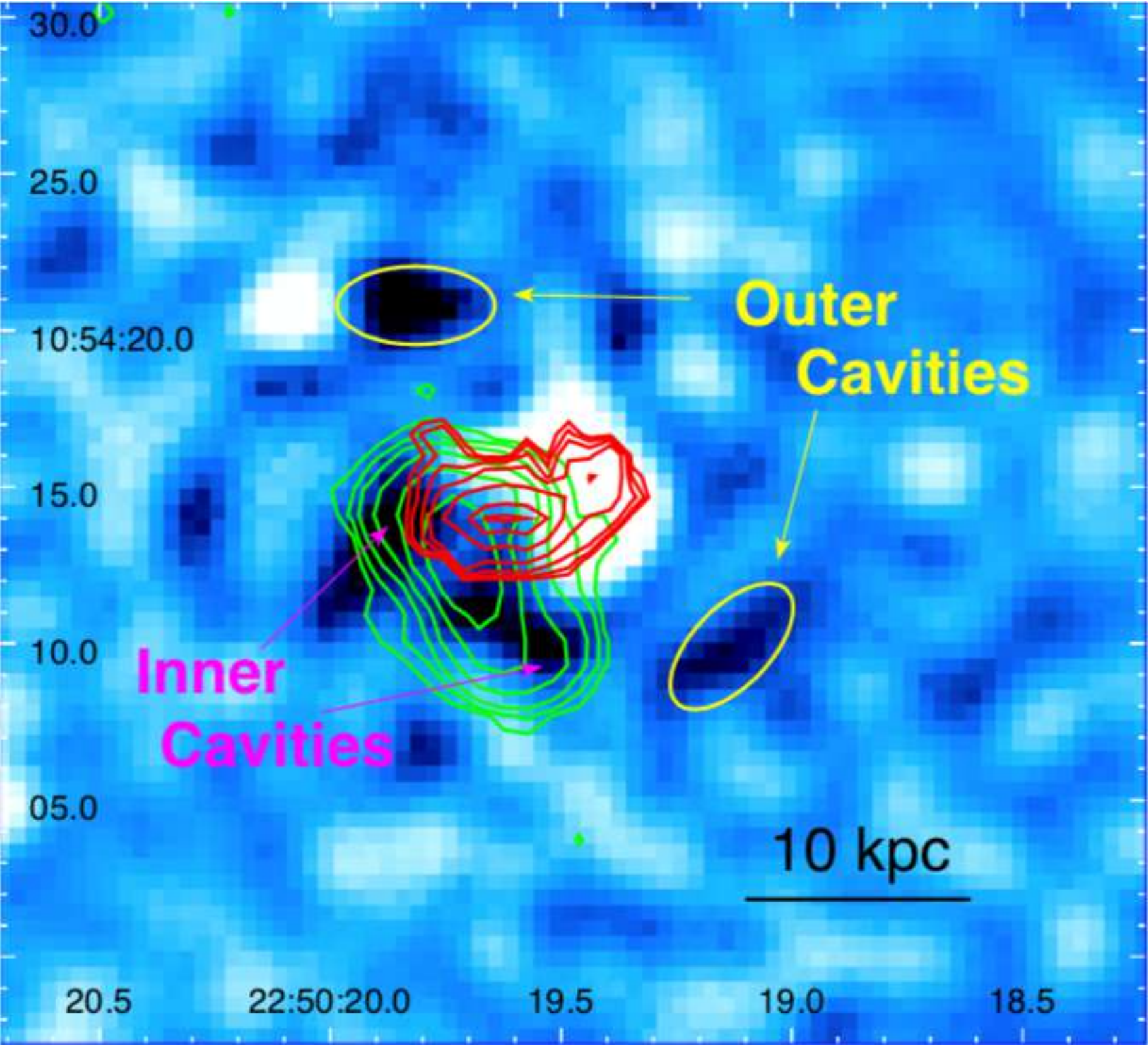}
	\caption{Residual image showing the two cavity systems located in A2495. The first pair is coincident with the radio lobes, while the second is symmetrical with respect to the X-ray peak region. The H$\alpha$ contours are overlaid in red, while the green contours are the same of Fig \ref{fig:multiwav}.}
	\label{fig:resid}
\end{figure}

We investigated the presence of cavities in A2495 by considering a residual image obtained by subtracting a 2-D $\beta$-Model from the 0.5-2 keV image. The result is shown in Fig. \ref{fig:resid}.

We notice an X-ray blob surrounding the peak region and four ICM depressions (two pairs), corresponding to $\sim$30$\%$ (at 90$\%$ confidence level) surface brightness deficits. Although we are aware of the limitations of our snapshot exposure, we notice that their shape and position, symmetrical with respect to the BCG for the first pair and to the X-ray peak for the second one(see Fig. \ref{fig:resid}), suggest that they could be real structures. In the following analysis and discussion, we assume them as cavities. However, it is possible that the poor statistic could have led to the production of artifacts; the reader shall thus be warned about the significance of these depressions. Deeper observations are therefore required to confirm their significance.
\\ \indent
The first pair seems to be coincident with the radio galaxy lobes, while the second pair is centered on, and falls on opposite sides of, the X-ray excess. Hereafter, we will refer to them as, respectively, \textit{inner} and \textit{outer} cavities. We also assume that cavities belonging to the same pair are characterised by the same elliptical shape and dimensions. Table \ref{tab:cav} lists their properties (\textit{a} refers to the major axis, \textit{b} to the minor one).

\begin{table}[!htb]
	\centering 
	\begin{tabular}{c c c c c}
		\hline
		\hline
		Cavity & a & b & V & R \\
		& [kpc] & [kpc] & [kpc$^3$] & [kpc]\\
		\hline
		Inner & 6.8 & 2.9 & 70.3 & 4.9 \\
		Outer & 8.0 & 4.9 & 164.0 & 11.9 \\
		\hline
	\end{tabular}
	\caption{Properties of the two cavities systems of A2495. a and b are the major and minor axis, V is the volume (estimated assuming an oblate elissoidal shape) while R is their distance from the BCG center.} \label{tab:cav}
\end{table}

\begin{figure}[h]
	\centering
	\includegraphics[width=26.5em]{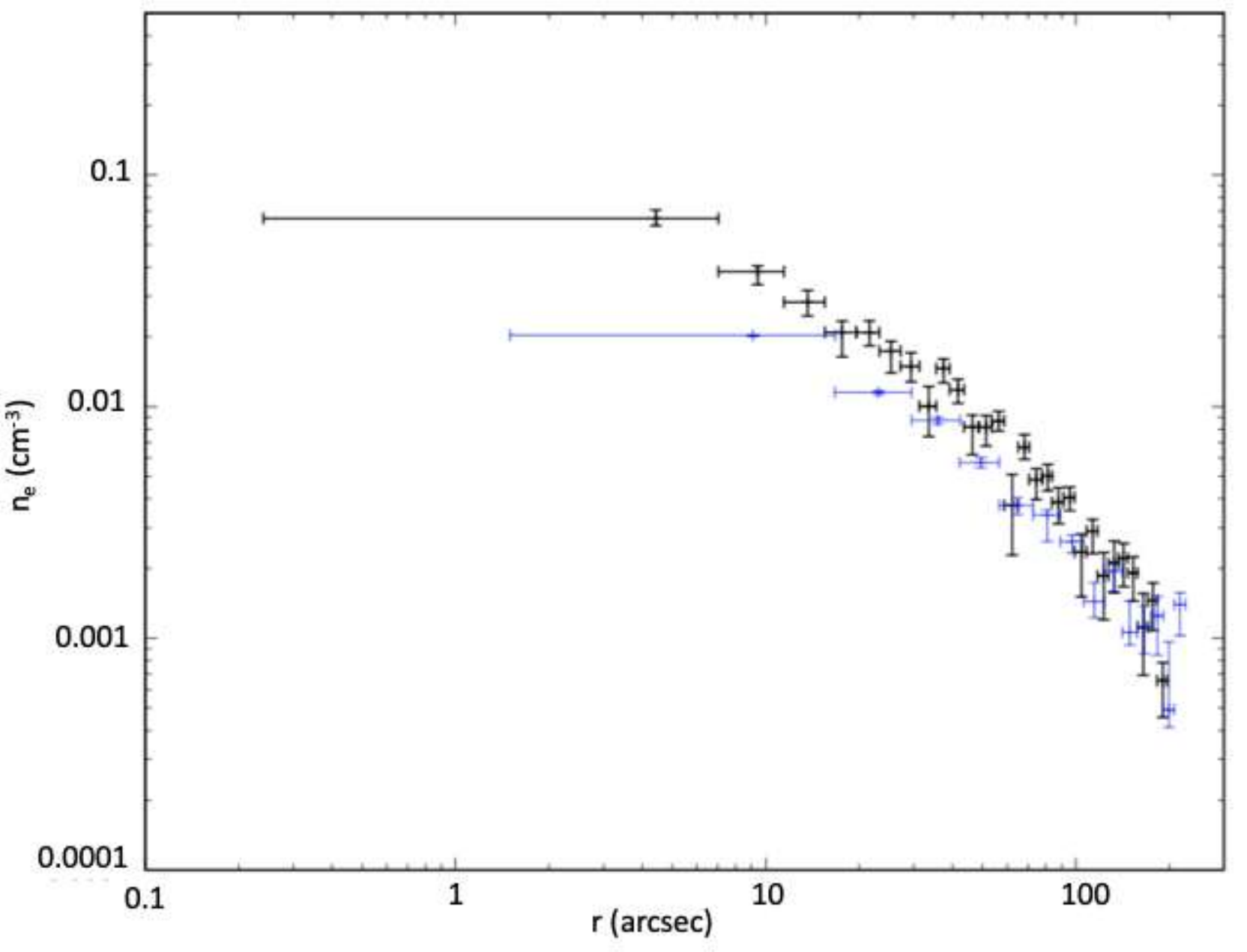}
	\caption{The black data are the results obtained from the double $\beta$-Model fit on the surface brightness profile, while the blue data are the results obtained from the spectral analysis described above.}
	\label{fig:confronto}
\end{figure}

Following the method described in \citet{Birzan_2004}, the cavity power can be estimated as:

\begin{equation}
P_{\text{cav}} = \dfrac{E_{\text{cav}}}{t_{\text{cav}}} = \dfrac{4pV}{t_{\text{cav}}}
\end{equation}
\label{pcav}

where $t_{\text{cav}}$ is the age of the cavity. 
In order to obtain an estimate of the temperature closer to the cavities' region, we performed a deprojection analysis with annular regions containing a minimum of 1500 counts; in fact, lowering the counts lower limit makes the rings radius smaller, bringing to a more localized estimate of the thermodynamical properties, albeit with larger uncertainties. However, we note that since we are close to the X-ray peak, the photon statistics remains high enough to obtain a good fit. We fitted a {\ttfamily project*wabs*apec} model and obtained the $kT$ value; results for the inner region, that corresponds to the radius we are interested in, are represented with a red dot in Fig. \ref{fig:temp}, \ref{fig:density}, \ref{fig:coolt} and \ref{fig:pressure}.
\\ \indent
Density was, instead, obtained exploiting the surface brightness profile showed in Fig. \ref{fig:surbri}. In fact, fitting a double $\beta$-Model on it \citep[for details of this method, see][]{Ettori_2002} allows us to obtain an estimate for the density values for smaller radii with respect to the spectral analysis performed above. Results are shown in Fig. \ref{fig:confronto}.

Since the two cavity systems lies at, respectively, $\sim$ 4.9 kpc and $\sim$ 11.9 kpc from the cluster center (see Table \ref{tab:cav}), we will make use of the first two density values: the first will be used for the inner cavities, the second for the outer ones. The \textit{kT} estimated above will instead be exploited for both the systems, since the current statistics does not allow us to provide more localized values.
In Table \ref{tab:1500} we finally report temperatures, densities and pressures used for the estimate of the cavities' properties.

\begin{table}[htb]
	\footnotesize
	\centering 
	\begin{tabular}{c c c c}
		\hline
		\hline
		$r_{min}$ - $r_{max}$ & kT$^{(a)}$ & $ n_e$$^{(b)}$ & p \\
		
		[arcsec]  & [keV] & [10$^{-2}$ cm$^{-3}$] & [10$^{-11}$ dy cm$^{-2}$] \\
		\hline
		0.2 - 7 & 2.37 $^{+0.64}_{-0.55}$ & 6.49 $^{+0.59}_{-0.45}$ & 45.0 $^{+14.3}_{-12.6}$ \\
		7 - 11.5 & 2.37 $^{+0.64}_{-0.55}$ & 3.83 $^{+0.22}_{-0.47}$ & 26.6 $^{+7.6}_{-8.9}$  \\
		\hline
	\end{tabular}
	\caption{Thermodynamical properties of the inner regions obtained from the deprojection analysis with 1500 counts ($\chi^2/$DoF $\sim$ 450/441) and from the fit of the surface brightness profile. The results of the spectral analysis are represented with red dots in Fig. \ref{fig:temp}, \ref{fig:density}, \ref{fig:coolt} and \ref{fig:pressure}.
	\\
    \textit{(a)} : obtained from the spectral analysis.
    \textit{(b)} : obtained from the surface brightness profile.
    } 
	\label{tab:1500}
\end{table}

\begin{table*}[]
	\begin{tabular}{ccccccc}
		&                       &        \textbf{Sound velocity}              &  &                       &                        \textbf{Buoyancy velocity}              &  \\ 
		\hline
		\hline
		\multicolumn{1}{c}{} & \multicolumn{1}{c}{$c_s \ [$km s$^{-1}]$} & \multicolumn{1}{c}{Age [Myr]} & \multicolumn{1}{c}{P$_{cav}$ [10$^{42}$ erg s$^{-1}$]} & \multicolumn{1}{c}{$v \ [$km s$^{-1}]$} & \multicolumn{1}{c}{Age [Myr]} & \multicolumn{1}{c}{P$_{cav}$ [10$^{42}$ erg s$^{-1}$]}  \\ 
		\hline
		\multicolumn{1}{c}{Inner} & \multicolumn{1}{c}{790 $^{+105}_{-83}$ } & \multicolumn{1}{c}{6.0 $^{+0.8}_{-0.6}$} & \multicolumn{1}{c}{18.3 $^{+7.5}_{-6.6}$} & \multicolumn{1}{c}{259 $\pm$ 58} & \multicolumn{1}{c}{18 $\pm$ 4} &  \multicolumn{1}{c}{6.0 $^{+2.3}_{-2.1}$}  \\ 
		\multicolumn{1}{c}{Outer} & \multicolumn{1}{c}{790 $^{+105}_{-83}$ } & \multicolumn{1}{c}{14.4 $^{+1.7}_{-1.5}$} & \multicolumn{1}{c}{10.4 $^{+3.7}_{-4.1}$} & \multicolumn{1}{c}{214 $\pm$ 42} & \multicolumn{1}{c}{53 $\pm$ 11} &  \multicolumn{1}{c}{2.8 $^{+0.9}_{-1.1}$ } \\ 
		\hline
	\end{tabular}
	\caption{Cavity velocity, age and power estimated with both the methods described in the text.}	\label{tab:results}
\end{table*}

The cavity age was calculated as $t_{\text{cav}} = R/v$, where \textit{R} is that defined in Table \ref{tab:cav}, while \textit{v} is the cavity velocity. In this work, we will consider the sound and the buoyancy velocity. 
The latter is defined as:

\begin{equation}
v_{\text{buoy}} = \sqrt{\dfrac{2gV}{CA}}
\end{equation}
\label{eq:vbuoy}

where $C$ $\simeq$ 0.75 represents the drag coefficient, \textit{V}
and \textit{A} are, respectively, the volume and the area of each
cavity, while \textit{g} is the gravitational acceleration.
\\ \indent
We estimated the latter by making use of the galaxy luminosity profile
(see Appendix) and of a simple model for the dark matter halo.
We adopted a stellar $M/L_\text{V}\sim 4$, appropriate for an old
stellar population with $\sim$ solar abundance \citep[es.][]{Maraston_2005}.
The resulting stellar masses within the region of the inner and outer
cavities are $M_{\text{BCG}}$($r \ <$ 4.9 kpc) $\sim
8.8 \cdot 10^{10}$ M$_\odot$ and $M_{\text{BCG}}$($r \ <$ 11.9 kpc)
$\sim 2.3 \cdot 10^{11}$ M$_\odot$, respectively. Given the
uncertainty on the mass-to-light ratio we adopted for these values a
30\% error.
\\ \indent
We estimated the dark matter contribution by assuming a NFW halo
\citep{Navarro_1996} of mass $M_{\text{NFW}} = 3.8 \cdot 10^{14}$
M$_\odot$, estimated from the $M-T$ relation of
\citealt{Finoguenov_2001}, and a concentration $c=4.65$
\citep{Prada_2012, Merten_2015}. We note that the assumed mass is
somewhat larger than the richness-based estimate by
\citealt{Andreon_2016} ($\sim 1.6 \cdot 10^{14} M_\sun$). On the other hand, it is consistent with the
hydrostatic mass \citep[e.g.,][]{Gitti_2012} determined exploiting the 
$\beta$-model (presented in Sec. \ref{sec:betamodel}), that returns $M \sim 3.2 \cdot 10^{14}$ M$_\odot$.
However, the conclusions for the cavity
dynamics discussed below are quite insensitive to the precise values
of $M_{\text{NFW}}$ and $c$. 
This model estimates dark masses of $M_{\text{DM}}(r < 4.9$ kpc$)=3.3 \cdot
10^{10}$ M$_\odot$ and $M_{\text{DM}}(r < 11.9$ kpc$)=1.8 \cdot 10^{11}$ M$_\odot$.
\\ \indent
We therefore obtain a total mass within the central 4.9 kpc of
$M_{\text{tot}}$($r \ <$ 4.9 kpc) $\sim 1.2 \cdot 10^{11}$ M$_\odot$,
while $M_{\text{tot}}$ ($r \ <$ 11.9 kpc) $\sim 4.1 \cdot 10^{11}$
M$_\odot$. Results of the cavity analysis are listed in Table \ref{tab:results}.
 \\ \indent
It is plausible that we are observing two different generations of cavities; the outer ones are older, while the inners probably still lie in the same regions they formed in. From this, we can infer that some process is probably switching the radio galaxy on and off: after the first generation forms, the AGN turns off and the cavities start to rise in the cluster atmosphere by buoyancy. Eventually, the AGN turns on again and another generation can be formed.
\\ \indent
Following the CCA, cold gas could originate if thermal instabilities ensue \citep[e.g.,][]{Gaspari_2012}. According to \citet{Voit_2015}, this is likely to happen when $t_{\text{cool}}/t_{\text{ff}} \leq 10-20$,
where $t_{\text{ff}} = \sqrt{2R^3/GM}$ is the \textit{free-fall time}. We note that, with the cooling time profile in Fig. \ref{fig:coolt} and the masses estimated above, this condition is not verified in A2495.
However, \citet{McNamara_2016} proposes that the $t_{\text{cool}}/t_{\text{ff}}$ ratio could be almost entirely governed by the cooling time. In both these interpretations, the cold gas could then accrete onto the SMBH efficiently and activate it.
\\ \indent
We propose that in A2495 the sloshing process mentioned above could, as well, play a part in the feedback cycle. When the cooler region approaches the BCG, the AGN begins to accrete and turns on, producing the first generation of cavities. Subsequently, because of sloshing, the accreting material diminishes and the SMBH is switched off. The oscillation could, at a later time, make the process to repeat. This would produce different generations of cavities.
\\ \indent
In order to investigate this scenario, we determined the free-fall time of the cooler region and compared it with the age difference of the two cavity generations, estimated with both the methods seen above. We obtained:

\begin{equation}
\normalsize
 \begin{cases}
t_{\text{ff}} \simeq 15.4 \pm 5.7 \ \text{Myr} \\ \Delta t_{\text{cav}}^{\text{sound}} \simeq 8.4\ ^{+2.1}_{-1.8} \ \text{Myr} \\ \Delta t_{\text{cav}}^{\text{buoy}} \simeq 35 \pm 12 \ \text{Myr}
 \end{cases}
 \end{equation}

We thus estimate $\Delta t_{\text{cav}} \simeq t_{\text{ff}}$ : the age difference between cavities reflects the time scale needed for the cooling region to move towards the BCG and activate the AGN. This is also consistent with the dynamical timescale obtained in Sec. \ref{sec:halpha}, that provides an alternative estimate for the offset lifetime.
\\ \indent
Future works, exploiting deeper observations, will likely be able to confirm or disprove our hypothesis.

\subsection{Can offset cooling  affect the feedback process?}

\begin{figure}[h]
	\centering
	\includegraphics[height=24.7em, width=26.3em]{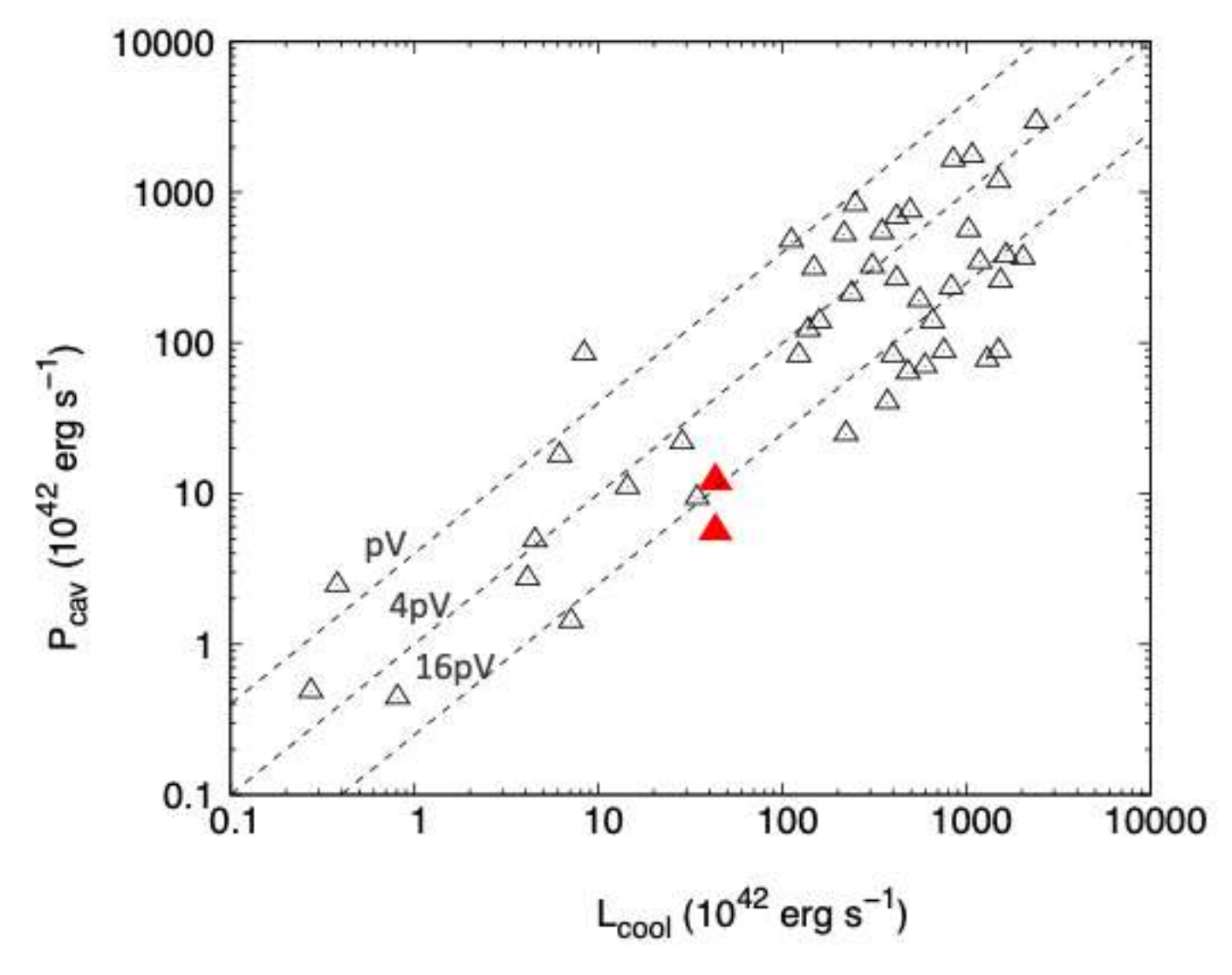}
	\caption{Black triangles represents the data from \citet{Birzan_2017}, while in red are the values for the two putative cavity systems considered in this work.}
	\label{fig:plotend}
\end{figure}

As we showed in the previous sections, the existing data suggest that cooling deposits gas away from the BCG nucleus so it cannot fuel the AGN. Therefore, we aim at investigating whether this can break the feeding-feedback cycle in A2495, or if the AGN activation cycle is driven by the periodicity of the gas sloshing motions.
\\ \indent
 We thus compared the power of the two putative cavity systems with the data available from samples of these structures \citep{Birzan_2017}. The result is presented in Fig. \ref{fig:plotend}, with the x axis representing the luminosity emitted within the cooling radius (estimated in Section \ref{sec:xanalisis}). For the cavity power, we used the values obtained from the buoyancy velocity, for consistency with the work cited above.

The values we estimated for A2495 are in good agreement with the global scatter of the observed relation. Therefore, we can argue that offset cooling seems to not break the feedback cyle, despite the evidence that cooling is not currently depositing gas into the BCG core where it can feed the AGN. Nevertheless, as we argued above, offsets probably still play a significant role in in the cycle regularization.

\section{Conclusions}

We carried out a thorough analysis of the X-ray and radio properties of A2495, by means of new observations requested with the purpose of performing a combined EVLA/\textit{Chandra} study of this cool-core galaxy cluster. We also discussed H$\alpha$ emission data from \citet{Hamer_2016} and exploited optical images retrieved from the HST archive. Our main results can be summarized as follows:
\begin{itemize}
\item The radio analysis at 1.4 and 5 GHz presented a small ($\sim$ 13-15 kpc) FRI radio galaxy with $L(1.4 \  \text{GHz}) \sim 2 \cdot 10^{23}$ W Hz$^{-1}$ and no apparent diffuse emission. This places A2495 as having the least radio powerful BCG among the 13 BCS objects that meet our selection criteria. The spectral index map highlighted a very steep synchrotron spectrum ($\alpha \simeq 1.35$), suggesting the presence of an old electronic population.
\item The X-ray study allowed us, through a deprojection analysis, to
  estimate the cooling radius of the cluster ($\sim$ 28$''$,
  corresponding to 40 kpc) and its luminosity (4.3 $\cdot 10^{43}$ erg
  s$^{-1}$).  Furthermore, we determined the cooling time, the density
  and the pressure radial profiles, finding $t_{\text{cool}} < 1$ Gyr inside
  $\sim 20$ kpc.
\item The multi-wavelength analysis showed two significant offsets, one of $\sim$ 6 kpc between the emission centroid and the X-ray peak, while the other one ($\sim$ 4 kpc) between the X-ray and the H$\alpha$ peaks. We propose that the first could be produced by sloshing of the ICM, while the second still remains unexplained and is worthy of more investigations. The line emitting plume connects the X-ray core emission to the BCG, suggesting that the origin of the 10$^4$ K gas could be linked to the cooling ICM. We found two putative cavity systems, the inner one with $t_{\text{age}} \sim 18$ Myr and $P_{\text{cav}} \sim$ 1.2 $\cdot$ 10$^{43}$ erg s$^{-1}$, while for the outer one $t_{\text{age}} \sim 53$ Myr and $P_{\text{cav}} \sim$ 5.6 $\cdot$ 10$^{42}$ erg s$^{-1}$. Their age difference is consistent with the free-fall time of the central cooling gas and with the offset dynamical timescale estimated from the line emitting gas; we thus suggest that the same sloshing motions could switch the AGN on and off, forming different generations of cavities.
\item Exploiting HST images, we located a dust lane crossing the BCG core in the North-South direction, and a $\sim$ 4$''$ filament extending West; we determined the dust mass within the central 7 kpc, finding $\sim$$1.7 \cdot 10^5$ M$_\sun$. We estimated in this region a lower limit for the total gas mass of $\sim$ 10$^7$ M$_\odot$, and argued that, since the H$\alpha$ structure can not entirely account for it, a significant fraction is still missing.  We propose that this fraction consists primarly of molecular gas; the $L_{\text{H}\alpha}$-$M_{\text{mol}}$ correlation supports this hypothesis, providing an estimate of $M_{\text{mol}} \sim 10^9$ M$_\sun$.
\item Finally, we proved that the offset cooling we found
does not break the feedback cycle, since the cavity power values for
A2495 with respect to the cooling luminosity are in agreement,
within the scatter, with the observed $P_{\text{cav}} - L_{\text{cool}}$ correlation. 
\end{itemize}

Deeper multi-wavelength observations (e.g., \textit{Chandra}, ALMA) will be required in order to better investigate the hypothesis we made in this work. A similar combined analysis of the other 12 BCS clusters which meet our selection criteria, and that apparently show similar features, will provide a better understanding of these dynamically active environments. 

\acknowledgments
We thank the referee for the careful reading of the manuscript and thoughtful comments and suggestions, that have significantly improved the presentation of our results. E.O'S. gratefully acknowledges funding the support for this work provided by the National Aeronautics and Space Administration (NASA) through \textit{Chandra} award number GO8-19112A issued by the \textit{Chandra} X-ray Center, which is operated by the Smithsonian Astrophysical Observatory for and on behalf of the National Aeronautics Space Administration under contract NAS8-03060.

\appendix
\label{appendix}
We exploited deep HST V-band observations to derive the
luminosity profile and structural parameters of the galaxy. The results obtained from this analysis have been
used in Section \ref{sec:cavities} to derive an estimate of the BCG mass.
As discussed by several authors \citep[e.g.,][]{Schombert_1987,
Graham_2005}, BCGs have a flux excess at large radii with respect to
the R$^{1/4}$ de Vaucouleurs law often used to fit the light profile
in ellipticals. Here we follow the approach outlined by
\citet{Donzelli_2011}, where the luminosity profile fitting procedure
uses a combination of a Sersic and an exponential function to
reproduce the inner and outer component of the surface brightness
profile. Briefly, the isophote fitting procedure ELLIPSE, within the
IRAF STSDAS package, was used to extract the luminosity profile from
the V-band image. Each function used in the light profile fitting
process provides a set of photometric parameters: from the Sersic model
we derived the surface brightness $\mu_e$ at $r=r_e$ (the half-light
radius), and the Sersic index $n$; the exponential model is
characterized by $\mu_0$ and $r_0$, which respectively represent the
surface brightness at the center and the length scale. Finally, the
total luminosity was computed by separately integrating the two fitting
models characterized by the photometric parameters.
\\ \indent
As a result of the fitting procedure, we obtained $r_e=11.9$ kpc and
$r_0=43.7$ kpc from the Sersic (best fit index = 1.89) and exponential
models, respectively. It is worth noting that a large fraction of the total
luminosity is contributed by the exponential model component, implying that a single de Vaucouleur or Sersic component could not reproduce the faint end of the light profile. The computed total luminosity is $L_\text{V}=4.1 \times 10^{11}$ L$_\sun$, which include a galactic extinction correction. In order to evaluate the galaxy mass at the location of inner and outer cavities (see Sec. \ref{sec:cavities}), we have computed the luminosities of $L_\text{V}$($r \ <$ 4.9 kpc) = 2.2 $\times 10^{10}$ L$_\sun$ and $L_\text{V}$($r \ <$ 11.9 kpc) = 5.8 $\times 10^{10}$ L$_\sun$.

\bibliographystyle{aasjournal}
\bibliography{bibliography}

\listofchanges
\end{document}